\documentclass[11pt]{article}
\pdfoutput=1
\usepackage{jheppub}

\usepackage{slashed}
\usepackage{xcolor}

\usepackage{hyperref}
\hypersetup{
    colorlinks=true,
    linkcolor=blue,
    filecolor=magenta,      
    urlcolor=cyan,
}
\def\lsim{\mathrel{\raise.3ex\hbox{$<$\kern-.75em\lower1ex\hbox{$\sim$}}}}
\def\gsim{\mathrel{\raise.3ex\hbox{$>$\kern-.75em\lower1ex\hbox{$\sim$}}}}

\definecolor{red}{rgb}{1.0, 0, 0}

\newcommand{\gev}{\text{GeV}}
\newcommand{\tev}{\text{TeV}}

\newcommand{\approxtext}[1]{\ensuremath{\stackrel{\text{#1}}{=}}}
\newcommand{\hatt}{\hat{\theta}}

\usepackage{jheppub} 
\usepackage{bm,float}
\usepackage[caption = false]{subfig}
\usepackage[section]{placeins} 
\usepackage{multirow, booktabs, amsthm, xfrac, fix-cm}
\usepackage{url, amsmath, amssymb, color, setspace, tabu}
\usepackage{tikz}
\usepackage{dsfont}

\usepackage{feynmp}
\DeclareMathOperator{\Tr}{Tr}

\usepackage{etoolbox}
\makeatletter
\patchcmd{\maketitle}{\@fpheader}{ }{ }{ }
\makeatother

\newcommand*{\mathcolor}{}
\def\mathcolor#1#{\mathcoloraux{#1}}
\newcommand*{\mathcoloraux}[3]{%
	\protect\leavevmode
	\begingroup
	\color#1{#2}#3%
	\endgroup
}

\newcommand{\Gtilde}{\widetilde{G}}
\newcommand{\Wtilde}{\widetilde{W}}
\newcommand{\Btilde}{\widetilde{B}}
\newcommand{\Hdag}{H^{\dag}}
\newcommand{\eps}{\epsilon}
\newcommand{\half}{\frac 1 2}

\newcommand{\dsix}{\text{dim-6}}
\newcommand{\deight}{\text{dim-8}}

\usepackage[utf8]{inputenc}

\title{On the impact of dimension-eight SMEFT operators on Higgs measurements}

\author[1]{Chris Hays}
\author[2]{, Adam Martin}
\author[3]{, Ver\'onica Sanz}
\author[3]{and Jack Setford}

\affiliation[1]{Department of Physics, University of Oxford, Oxford OX1 3RH, UK}
\affiliation[2]{Department of Physics, University of Notre Dame, Notre Dame, IN, 46556, USA}
\affiliation[3]{Department of Physics and Astronomy, University of Sussex, Brighton BN1 9QH, UK}

\date{\today}
\preprint{ }

\abstract{Using the production of a Higgs boson in association with a $W$ boson as a test case, we assess the impact of 
dimension-8 operators within the context of the Standard Model Effective Field Theory.  Dimension-8--SM-interference and 
dimension-6-squared terms appear at the same order in an expansion in $1/\Lambda$, hence dimension-8 effects can be treated 
as a systematic uncertainty on the new physics inferred from analyses using dimension-6 operators alone. To study the 
phenomenological consequences of dimension-8 operators, one must first determine the complete set of operators that can 
contribute to a given process. We accomplish this through a combination of Hilbert series methods, which yield the number of 
invariants and their field content, and a step-by-step recipe to convert the Hilbert series output into a phenomenologically 
useful format. The recipe we provide is general and applies to any other process within the dimension $\le 8$ Standard Model Effective Theory. We quantify the effects of dimension-8 by turning on one dimension-6 operator at a time and setting all dimension-8 operator coefficients to the same magnitude. Under this procedure and given the current accuracy on $\sigma(pp \to h\,W^+)$, we find the effect of 
dimension-8 operators on the inferred new physics scale to be small, $\mathcal O(\text{few}\,\%)$, with some variation depending on the relative signs of the dimension-8 coefficients and on which 
dimension-6 operator is considered. The impact of the dimension-8 terms grows as $\sigma(pp \to h\,W^+)$ is measured more accurately 
or (more significantly) in high-mass kinematic regions. We provide a FeynRules implementation of our operator set to be used 
for further more detailed analyses.
}
\emailAdd{chris.hays@physics.ox.ac.uk}
\emailAdd{amarti41@nd.edu}
\emailAdd{v.sanz@sussex.ac.uk}
\emailAdd{j.setford@sussex.ac.uk}

\keywords{}

\begin{document}
\raggedbottom
\maketitle
\flushbottom

\section{Introduction and motivation}
\label{sec:intro}

The particle physicist's dream---direct detection of high-scale physics beyond the Standard Model (BSM)---has yet to be 
realised. As we transition into an era of precision Higgs measurements, an appropriate framework for testing the Standard Model 
(SM) is to treat it as an Effective Field Theory (EFT). From an EFT perspective, the SM Lagrangian is merely the first few terms 
in the (infinite) series:
\begin{equation}
\mathcal L = \sum_d \sum_i \frac{c_d^{(i)}}{\Lambda^{4-d}}\,\mathcal O_d^{(i)},
\label{eq:EFT}
\end{equation}
where $\Lambda$ is the cutoff of the effective theory, $c_d^{(i)}$ are the Wilson coefficients, and $\mathcal O_d^{(i)}$ are 
all the gauge-invariant operators at mass-dimension $d$ involving the Standard Model fields.

The direct observation of new physics might be beyond the reach of the LHC, but it could still manifest \emph{indirectly} as 
contributions to the Wilson coefficients of the effective theory.  Constraining these coefficients is a powerful and general 
way to probe BSM models, since any weakly-coupled high-mass state can be integrated out to give a particular pattern of 
coefficient values.

The current state of the art is the classification and study of dimension-6 operators\,\cite{Weinberg:1979sa, Buchmuller:1985jz, 
Lehman:2014jma, Henning:2015alf}. Several different non-redundant bases have been constructed\,\cite{Grzadkowski:2010es, 
Contino:2013kra, Gupta:2014rxa, Masso:2014xra, deFlorian:2016spz}, as have dictionaries that allow translation between 
them\,\cite{Falkowski:2015wza, Aebischer:2017ugx}. Via detailed phenomenological studies and global fits, many of the 
Wilson coefficients of these operators have been constrained\,\cite{Han:2004az, Pomarol:2013zra, Ellis:2014dva, Ellis:2014jta, 
Murphy:2017omb, Falkowski:2014tna, Efrati:2015eaa, Falkowski:2015jaa, Falkowski:2015krw, Falkowski:2016cxu, Falkowski:2017pss, 
Falkowski:2018dmy, Corbett:2012ja, Corbett:2013pja, Englert:2015hrx, Buckley:2016cfg, Buckley:2015lku, Corbett:2015ksa, 
Butter:2016cvz, Dumont:2013wma, deBlas:2017wmn, deBlas:2016nqo, Berthier:2015oma, Berthier:2015gja, Berthier:2016tkq, 
Brivio:2017bnu,Farina:2016rws,Banerjee:2018bio}. Some progress has also be made at dimension-8, particularly in the gauge sector\,\cite{Degrande:2013kka,Liu:2016idz, Liu:2018pkg}.

Representing the scale at which the new physics appears as $\Lambda$, one can schematically write the amplitude for a given 
process as
\begin{equation}
\mathcal A \sim \left( A_{\textrm{SM}} + \frac{A_{\textrm{\dsix}}}{\Lambda^2}+ \dots \right),
\end{equation}
where $A_{\textrm{SM}}$ stands for the Standard Model amplitude, and $A_{\textrm{\dsix}}$ is the correction coming from the 
dimension-6 EFT. The leading-order correction to the cross-section, at $\mathcal O(1/\Lambda^2)$, is therefore an interference 
term of the form $A_{\textrm{SM}} \times A_{\textrm{\dsix}}$.  We also have an $|A_{\textrm{\dsix}}|^2$ correction at order 
$\mathcal O(1/\Lambda^4)$, which can be readily computed with current dimension-6 technology. However, consistent power-counting 
would require that, alongside $|A_{\textrm{\dsix}}|^2$ terms, we include $A_{\textrm{SM}} \times A_{\textrm{\deight}}$ 
interference effects, where now:
\begin{equation}
\mathcal A \sim \left(A_{\textrm{SM}} +\frac{A_{\textrm{\dsix}}}{\Lambda^2} +\frac{A_{\textrm{\deight}}}{\Lambda^4} +\dots\right).
\label{eq:powercounting}
\end{equation}
In other words, $A_{\textrm{SM}}\times A_{\textrm{\deight}}$ interference and $|A_{\textrm{\dsix}}|^2$ terms both appear at 
$\mathcal O(1/\Lambda^4)$, and so naively should be given equal consideration. Given that so much progress has already been 
made at dimension-6, it is natural to ask whether complementing our analyses with dimension-8 operators can have any 
effect on current constraints.

The are two distinct scenarios one might consider: 

\begin{enumerate}
\item The new physics is dominated by $A_\textrm{SM}\times A_{\textrm{\dsix}}$, i.e. interference terms between 
dimension-6 contributions and the Standard Model.  This is what one would naively expect whenever the scale of new physics 
is high in comparison with the electroweak scale. 
\item The leading effect of new physics is instead $|A_{\textrm{\dsix}}|^2$, potentially of the same order as the 
$A_{\textrm{SM}}\times A_{\textrm{\deight}}$ interference terms.  This situation could arise due an accidental suppression, 
or indicate an underlying structure in the BSM context.  For example, the interference could be suppressed by ratios of the 
weak scale to the cutoff $\Lambda$ to some power.  One way this can occur is if there is a helicity mismatch between the SM 
and dimension-6 operators\,\cite{Azatov:2016sqh}. Another possibility is if the dimension-6 operators are purely CP-odd and 
the observable is a CP-even quantity.  Or, when focusing on specific kinematic regimes, the SM content could be small and 
lead to the dominance of $|A_{\textrm{\dsix}}|^2$ effects on a particular experimental bin.
\end{enumerate}

In each of these cases, we would like to know whether including dimension-8 operators can significantly alter the bounds we 
put on the dimension-6 Wilson coefficients. It may prove that an analysis purely at the dimension-6 level is inadequate, or 
that dimension-8 effects should be accounted for as a systematic uncertainty.

Constructing a complete set of effective operators at any given order is not a trivial task.  It is well known that including 
every operator allowed by the symmetries of the theory results in an over-complete set: operators will be related to others 
via the equations of motion (EOM) and/or integration by parts (IBP).  As discussed and illustrated in 
Refs.\,\cite{Lehman:2015via, Lehman:2015coa}, the task is facilitated by a mathematical tool known as the Hilbert series.  
Given a set of objects transforming in representations of the symmetry group, the Hilbert series generates all polynomials of 
the objects---incorporating the symmetry (or antisymmetry) under the interchange of identical objects---and projects out all 
invariants using character orthonormality.  Hilbert series techniques have found use in a number of 
theoretical\,\cite{Pouliot:1998yv, Benvenuti:2006qr, Dolan:2007rq, Gray:2008yu, Hanany:2008sb, Chen:2011wn, Butti:2007jv, 
Feng:2007ur, Forcella:2007wk, Benvenuti:2010pq, Hanany:2012dm, Rodriguez-Gomez:2013dpa, Dey:2013fea, Hanany:2014hia, 
Begin:1998hn, Hanany:2014dia, Henning:2017fpj} and phenomenological\,\cite{Jenkins:2009dy, Hanany:2010vu, Merle:2011vy, 
Lehman:2015via, Lehman:2015coa, Henning:2015alf, Kobach:2017xkw,Kobach:2018nmt} contexts.  Applied to the SMEFT, the relevant 
symmetry group for the Hilbert series is the Lorentz symmetry group plus the $SU(3) \otimes SU(2) \otimes U(1)$ gauge groups, 
and the objects entering the Hilbert series are the SM fields $\{ Q,u^c, d^c, L, e^c, H\}$ plus the gauge field strengths.  
As character orthonormality only strictly applies to compact groups, we work with the Euclideanized Lorentz symmetry 
$SO(4) \cong SU(2)_L \otimes SU(2)_R$.  Furthermore, to correctly incorporate IBP redundancies we place all objects into 
representations of the conformal $SO(4,2)$ group instead of the Lorentz group.  The conformal representations package a field 
and all of its derivatives into a single object\footnote{By placing scalars, fermions, and field strengths into {\em short} 
representations of $SO(4,2)$ we can remove equation of motion (EOM) redundancies.}, and we remove the IBP redundancy from any 
product of objects by projecting out the highest weight component (again via character orthonormality, with some minor 
modifications due to the non-compact nature of $SO(4,2)$).

In the SMEFT it is natural to order the invariants by their mass dimension.  The output of the Hilbert series is the number 
of invariants at each mass dimension, and the field content of the invariants.  For example, applying the Hilbert series to 
the SMEFT, one of the dimension-6 invariants is
\begin{align}
2 D^2(\Hdag H)^2,
\label{eq:d6example}
\end{align}
indicating that there are two operators containing two derivatives and four Higgs fields.  What the Hilbert series does 
\emph{not} tell us is how the various indices carried by each of the fields should be contracted.  We must manually translate 
the output of the Hilbert series into a format that is useful for calculating Feynman rules, or for inputting into 
computational tools such as FeynRules~\cite{Christensen:2008py, Degrande:2011ua, Alloul:2013bka}.  In the case of the operator 
in Eq.~\eqref{eq:d6example}, we need to work out which fields the derivatives act on and how the Lorentz and $SU(2)_w$ indices 
are contracted.

Converting the Hilbert series output into a more familiar form is one of the goals of this paper.  The full set of 993 
dimension-8 SMEFT operators\footnote{The value of 993 holds for $N_f = 1$, where $N_f$ is the number of generations, counting 
operators and their hermitian conjugates as separate operators.  For three generations the number of dimension-8 SMEFT operators 
jumps to 44807~\cite{Henning:2015alf}.  See Sec.~\ref{sec:D0} for more information.} is listed in Ref.\,\cite{Henning:2015alf} 
and available in Mathematica format.  While the recipe we outline in the following will work for all operators of any mass 
dimension, in this paper we explicitly work out the conversion from Hilbert series output to operators for only a subset of 
dimension-8 operators.  It is our hope that the recipe and examples provided will facilitate the extraction of the remaining 
operators for the purposes of more general and in-depth analyses.  Looking to phenomenology for guidance, we have chosen to 
illustrate the procedure for operators which contribute to $pp \to h\, W$.  This process is a good choice for several reasons.  
First, it is a relatively clean process that can be measured at high $Q^2$, where higher dimensional operators become 
important~\cite{Isidori:2013cla, Isidori:2013cga}.  Second, it severely restricts the number of relevant dimension-8 operators.  
After translating the set of required operators, we investigate $pp \to h\, W$, exploring how constraints on 
dimension-6 Wilson coefficients are affected by the inclusion of dimension-8 effects.  Our analysis focusses on one specific 
example but in principle one can use this technology to study many other processes, for example a Higgs boson produced in 
association with top quarks, or through vector boson fusion.  To this end we also make publicly available the FeynRules 
implementation of this set of operators, so that future work can extend our analysis.

\section{Dimension-8 operator set}

The SMEFT inputs to the Hilbert series are the matter fields $\{Q, u^c, d^c, L, e^c, H\}$, their hermitian conjugates, and the 
field strength tensors.  We take all fermions to be left-handed and work with combinations of the field strengths and their duals 
$X_{L,R}^{\mu\nu} = \frac{1}{2}(X^{\mu\nu} \mp i \widetilde{X}^{\mu\nu})$, where $X = B, W, G$ for hypercharge, $SU(2)_w$ and 
$SU(3)_c$ respectively, since they have simpler Lorentz transformation properties ($X_{L} \sim (1,0),~X_R \sim (0,1)$ under 
$SU(2)_L \otimes SU(2)_R$).  We list the full set of SMEFT representations in Table~\ref{tbl:representations} using the 
convention $(SU(2)_L$, $SU(2)_R$;~$SU(3)_c$,~$SU(2)_w$,~$U(1)_Y)$.
\begin{table}[h!]
\centering
\begin{tabular}{c|cc|c}
$Q$ & $(\half, 0; 3, \half, \frac 1 6)$ & $H$ & $(0, 0; 0, \half, \half)$ \\
$u^c$ & $(\half, 0; \bar 3, 0, -\frac 2 3)$ & $B_L$ & $(1,0;0,0,0)$ \\
$d^c$ & $(\half, 0; \bar 3, 0, \frac 1 3)$ & $W_L$ & $(1,0;0,1,0)$ \\
$L$ & $(\half, 0; 0, \half, -\frac 1 2)$ & $G_L$ & $(1,0;8,0,0)$ \\
$e^c$ & $(\half, 0; 0, 0, 1)$ \\
\end{tabular}
\caption{Representations used in the Hilbert series construction.  Hermitian conjugate representations have 
$SU(2)_L \leftrightarrow SU(2)_R$ and all gauge representations replaced by their conjugates (when applicable). }
\label{tbl:representations}
\end{table}
The fields are dressed with characters corresponding to their gauge and {\em conformal} representations, and 
plugged into the Hilbert series generating function.  Details of the characters, and of the reduction of the 
polynomial produced by the generating function to gauge and Lorentz singlets, can be found in 
Refs.~\cite{Lehman:2015via, Lehman:2015coa,Henning:2015alf}.  Weighting each operator in the output by its 
mass dimension, we can easily filter out the dimension-8 operators. 

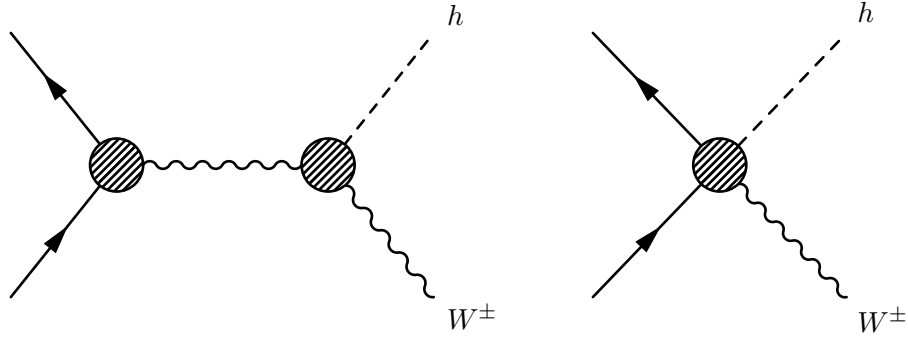
\begin{figure}[tb!]
\begin{fmffile}{pphW_feynman}
\begin{equation*}
\label{pphW_feynman}
\begin{tikzpicture}[baseline=(current bounding box.center)]
\node{
\fmfframe(1,1)(1,1){
\begin{fmfgraph*}(200,100)
\fmfleft{i1,i2}
\fmfright{f1,f2}
\fmf{fermion}{i1,v1}
\fmf{fermion}{v1,i2}
\fmf{dashes}{v2,f2}
\fmf{wiggly}{v2,f1}
\fmf{wiggly}{v1,v2}
\fmflabel{$W^\pm$}{f1}
\fmflabel{$h$}{f2}
\fmfblob{20}{v1}
\fmfblob{20}{v2}
\end{fmfgraph*}	
}	
};
\end{tikzpicture}
\;\;\;\;\;
\begin{tikzpicture}[baseline=(current bounding box.center)]
\node{
\fmfframe(1,1)(1,1){
\begin{fmfgraph*}(120,100)
\fmfleft{i1,i2}
\fmfright{f1,f2}
\fmf{fermion}{i1,v1}
\fmf{fermion}{v1,i2}
\fmf{dashes}{v1,f2}
\fmf{wiggly}{v1,f1}
\fmflabel{$W^\pm$}{f1}
\fmflabel{$h$}{f2}
\fmfblob{20}{v1}
\end{fmfgraph*}
}
};
\end{tikzpicture}
\end{equation*}
\end{fmffile}
\caption{Diagrams contributing to $pp\rightarrow h\,W^\pm$.  The shaded circles represent integrated-out new physics, and mark 
the vertices modified by the HDOs we have included in our analysis. \emph{Left}: Associated production via an $s$-channel $W$ 
boson.  In this case the purely bosonic operators modify the $hWW$ coupling, and the contact operators modify the quark-$W$ 
vertex. \emph{Right}: The four-point contact interaction that can also mediate $pp\rightarrow h\,W^\pm$.}
\label{diagrams}
\end{figure}

As mentioned in the introduction, we have chosen to illustrate the procedure for operators that can contribute to 
$pp \to h\, W$ processes.  Following the power-counting argument in Eq.~\eqref{eq:powercounting}, we are interested in the 
subset of dimension-8 operators that can {\em interfere} with the SM contributions to $pp \to h\, W$.  If we take all first 
and second generation fermions to be massless\footnote{Also ignoring bottom and top quark components of the proton.}, the 
leading-order SM $pp \to h\, W$ amplitudes all proceed through a single topology: the process is initiated by a left-handed 
quark and an antiquark that annihilate into a $W$ boson, which then emits a Higgs boson.  The requirement that dimension-8 
operators interfere with this amplitude limits us to three types of operator\footnote{Here we are referring to new vertices. 
Dimension-8 operators can also indirectly enter $pp \to h\,W^+$ through field redefinitions or through the relations between 
couplings and experimental inputs. We will study these effects in more detail in Sec.~\ref{application} and 
Appendix~\ref{app:redefs}.}: 1.) operators that modify the $hWW$ vertex, 
2.) operators that modify the $\bar q q W$ vertex, and 3.) four-point contact operators.  The relevant Feynman diagrams are 
shown in Figure~\ref{diagrams}.  Operators of the first type are purely bosonic and involve Higgs fields, derivatives, and 
field strengths, while operators of the latter two types must involve a like-chirality fermion-antifermion pair, at most one 
field strength, Higgs fields and derivatives.  All three operator types must be included in order to have a basis-independent 
result.  We will not consider modifications to the $W$-boson couplings to leptons; in doing so we are assuming that the $W$ 
boson is an on-shell final-state particle.

We now extract from the Hilbert series the relevant pieces for the operators we wish to consider.  Due to the way EOM are handled 
in the Hilbert series machinery, its output is always in the so-called Warsaw basis~\cite{Grzadkowski:2010es}, where higher 
derivative terms are removed in favor of operators with more fields whenever possible.  Therefore, when combining dimension-8 
and dimension-6 effects in later sections, we use the Warsaw basis for dimension-6.  For other advantages of the Warsaw basis, 
see Refs.\,\cite{Alonso:2013hga,Azatov:2016sqh, Brivio:2017btx}.  Focusing first on the bosonic operators, we can group the 
dimension-8 operators according to the number of derivatives. At $\mathcal O(D^0)$, we find:
\begin{equation}
\begin{gathered}
(H^{\dag}H)^4, \enskip
(H^{\dag}H)^2(B_L)^2, \enskip
(H^{\dag}H)^2\,B_L W_L, \enskip
2\,(H^{\dag}H)^2\,(W_L)^2, \\
(H^{\dag}H)^2(G_L)^2, \enskip
(H^{\dag}H) B_L (W_L)^2, \enskip
(H^{\dag}H) (W_L)^3, \enskip
(H^{\dag}H) (G_L)^3,
\end{gathered}
\label{eq:HS1}
\end{equation}
where for all operators except $(H^{\dag}H)^4$ there is a corresponding hermitian conjugate operator with 
$B_L, W_L, G_L \rightarrow B_R, W_R, G_R$.  At $\mathcal O(D^2)$ the operators are:
\begin{equation}
\begin{gathered}
 2\,D^2 (H^{\dag}H)^3, \enskip
 D^2 (H^{\dag}H\,B_L B_R), \enskip
 D^2 (H^{\dag}H\, G_L G_R), \enskip
 2\, D^2 (H^{\dag}H\, W_L W_R) \\
 D^2 (H^\dag H)^2 B_L, \enskip
 D^2 (H^{\dag}H\,(B_L)^2), \enskip
 D^2 (H^{\dag}H\,(G_L)^2),  \enskip
 2\, D^2 (H^{\dag}H\,(W_L)^2), \\
 2\, D^2 (H^{\dag}H\,B_L W_L), \enskip
 D^2 (H^{\dag}H\, B_R W_L), \enskip
 2\, D^2 ((H^{\dag}H)^2 W_L).
\end{gathered}
\label{eq:HS2}
\end{equation} 
The first row of operators are self-hermitian, while hermitian conjugates must be added for the operators 
in the second and third rows.  Finally, at $\mathcal O(D^4)$ there is one operator set:
\begin{equation}
3\, D^4 (H^2\, H^{\dag2}),
\end{equation}
which is self-hermitian.  Moving on to the contact operators and operators that modify the $\bar q q\, W$ vertices, 
the Hilbert series output for left-handed quarks is:
\begin{equation}
\begin{gathered}
4\, D(Q^{\dag}Q\, (H^{\dag}H)^2), 6\, D(Q^{\dag}Q\, H^{\dag}H\, W_{L}), \\ 
4\, D^3(Q^{\dag}Q\, H^{\dag}H),
\end{gathered}
\label{eq:HS4}
\end{equation}
with additional hermitian conjugates for the $W_L$ terms.  The operators with no field strengths will impact the 
$\bar q q W$ couplings, and all sets will generate $\bar q q W h$ contact terms.  We could have written the fermionic 
operators with a generation index, leading to a different operator for each possible generation combination, 
$Q^{\dag}_1 Q_3, Q^{\dag}_2 Q_1$, etc. In some circumstances, adding generation indices would disrupt the antisymmetrization 
we must perform when an operator contains multiple identical fermionic fields.  For these operators, there are no repeated 
fields, so adding generation indices would just multiply the number of operators by $N_f^2$.  Throughout this paper we will 
ignore this complication and assume the couplings are universal among generations.

As we can see, the Hilbert series tells us exactly how many independent invariants we can construct from each combination of 
fields, but it does not tell us how the indices should be contracted.  In the next two subsections we detail how to convert this 
output into canonical phenomenological form.  Readers more interested in the applications of dimension-8 operators can skip to 
Sec.~\ref{sec:bos8}.

\subsection{Operators with zero or one derivative}
\label{sec:D0}

For operators with zero or one derivative, the process of converting the Hilbert series output into useful phenomenological 
form is fairly straightforward and can be broken down into six steps:

\begin{enumerate}
\item if the operator contains multiple instances of the same field ($H$, $Q$, etc.), work out the group products that are 
allowed given the Bose or Fermi symmetry;
\item from the properly symmetrized blocks, determine the contractions that lead to an overall invariant for the entire operator; 
\item express the contractions using the available group-theory objects, e.g. $\epsilon_{ab}$ for $SU(2)$ and $\delta^A_B$ or 
$f_{ABC}$ for $SU(3)$;
\item translate Lorentz contractions from $SU(2)_L \otimes SU(2)_R$ to $SO(3,1)$;
\item translate from the L and R field-strength combinations to the field strength $X_{\mu\nu}$ and its dual $\tilde X_{\mu\nu}$;
\item impose hermiticity.
\end{enumerate}

Steps 1 and 2 require picking out singlets from products of representations, thereby dictating how the various indices need to 
be contracted, while steps 3 through 5 take us from the formalism convenient for group theory to canonical operator conventions.  
We emphasize that the six steps give us the form of the operators corresponding to the Hilbert series and {\em not} the 
coefficient accompanying the operator, i.e. the factor $c^{(i)}_d$ in Eq.~\eqref{eq:EFT}.  The coefficients can only be set 
knowing the UV theory.  Because of this ignorance of the operator coefficient, we will completely ignore all overall numerical 
(and sign) factors that appear when translating operators, subsuming them into the $c^{(i)}_d$. 

The conversion steps are best illustrated with an example.  We will work out the steps for this example by hand, though steps 
1 and 2 can also be performed using programs such as {\tt susyno}\,\cite{Fonseca:2011sy}.  While it is often obvious to see how 
the indices on an operator must be contracted without plodding through every step, we will go through this example in full detail 
so that the process can be applied in cases where inspection fails.  Other examples can be found in Appendix~\ref{app:examples}.

\subsubsection{Example: $(H^{\dag}H)^2 B^2_L$}
\label{sec:example1}

Before moving on to the operator conversion, let us define our index conventions.  Most of the index manipulations will involve 
$SU(2)$.  We will use undotted Greek letters for $SU(2)_L$ indices, dotted Greek letters for $SU(2)_R$ indices, and Latin letters 
for $SU(2)_w$.  When we convert from Lorentz $SU(2)_L \otimes SU(2)_R$ to $SO(3,1)$ we will also use Greek indices for $SO(3,1)$, 
though drawn from the middle of the alphabet (which should be clear from the context).  Symmetrized groups of indices will be 
indicated by curly braces $\{\alpha\beta\}$ and antisymmetric ones by brackets, $[ab]$.  Occasionally we will find it convenient 
to convert products of $SU(2)$ doublets into triplets using Pauli matrices, which will be indicated by upper-case Latin letters 
(i.e. $\tau^I$).  While $SU(2)$ doublet indices can only be contracted by $\epsilon_{ij}$\footnote{We define 
$\epsilon^{12} = -\epsilon^{21} = \epsilon_{21} = -\epsilon_{12} = 1$, following the conventions in \cite{Dreiner:2008tw}. }, 
triplet indices may be contracted by $\delta^{IJ}$ or the antisymmetric $\epsilon_{IJK}$.  As an example, the singlet contraction 
of the $SU(2)$ doublets $\Hdag$ and $H$ is given by $\epsilon^{ij} H^\dagger_i H_j$.  The form $\epsilon^{ij} H^\dagger_i H_j$ 
may seem a bit strange at first glance. It occurs because we are working with $H^{\dag}$ (and $L^{\dag}, Q^{\dag}$) as a $2$ of 
$SU(2)$ rather than a $\bar 2$, as this makes the group manipulation simpler.  One can easily convert between the $2$ and 
$\bar 2$ forms via $\epsilon$:  $H^{\dag}_{\bar 2} = \epsilon \Hdag_{2}$. Throughout this paper we will perform all intermediate 
steps using $\Hdag_2$, then convert to the more familiar $\Hdag_{\bar 2}$ form at the end.

Now we turn to our first example: $(H^\dagger H)^2 B_L^2$.  As the coefficient of this Hilbert series output is 1, there is one 
invariant to find.  The operator involves three different fields, each of which occurs twice.  As $H^{\dag} \ne H$ and $B_L$ are 
all bosonic, each block of $(\text{field})^2$ must be symmetric.  Starting with the Higgs field and dropping the $SU(3)_c$ and 
$U(1)_Y$ entries for brevity, 
\begin{align}
H^2 = (0,0;\frac 1 2)^2_{symm} = (0,0; 0 \oplus 1)_{symm} = (0,0; 1).
\label{eq:H21}
\end{align}
This is telling us nothing more than $H_i\, H_j$ contracted with $\epsilon^{ij}$ gives zero (where $i,j$ are $SU(2)_w$ indices). 
Only the symmetric combination $H_{\{i,}H_{j\}}$, or $H^2_{\{ij\}}$, is nonzero. $(H^{\dag})^2$ is the same as $H^2$, so we can 
proceed to $B^2_L$:
\begin{align}
B^2_L = (1,0; 0)^2_{symm} = (0 \oplus 1 \oplus 2,0;0)_{symm} = (0\oplus 2, 0; 0).
\end{align}
This decomposition is also straightforward: the symmetric product of two spin-1 fields must be spin-0 or spin-2\footnote{The 
reader may wonder why we are using Lorentz representations in this section after alluding to conformal representations in 
Sec.~\ref{sec:intro}.  Conformal representations package an object and all its derivatives together and are useful for 
automatically incorporating IBP and EOM redundancies.  To explicitly construct operators for given field (and derivative) 
content, working with Lorentz representations is sufficient.}.  Multiplying the three blocks together gives:
\begin{align}
H^2 (H^{\dag})^2 B^2_L = (0,0; 1) \otimes (0,0; 1) \otimes (0 \oplus 2,0;0) = (0 \oplus 2, 0; 0 \oplus 1 \oplus 2).
\end{align}
Choosing the $0$ from the $SU(2)_L$ and $SU(2)_w$ products we get the one invariant promised by the Hilbert series.  Notice 
we have $0 \oplus 1 \oplus 2$ in the $SU(2)_w$ portion since there is no (anti)symmetrization left to eliminate a piece. 

The above procedure takes us through step 2.  For step 3, we can focus on the Higgs portion since that contains all the 
$SU(2)_w$ dependence.  For $SU(2)$, the only object available to contract indices is $\epsilon^{ij}$.  The contraction that 
does not vanish is:
\begin{align}
H^2_{\{ij\}} (H^{\dag})^2_{\{km\}}\epsilon^{ik}\epsilon^{jm}.
\end{align}
The contraction of $i$ with $m$ and $j$ with $k$ is equivalent since $i \leftrightarrow j$, $k \leftrightarrow m$ is symmetric, 
and contracting $i$ with $j$ gives zero.  Written in a more familiar way, we have
\begin{align}
(\epsilon H^{\dag}H)^2
\end{align}
where we have ignored any overall factors of 2 or -1.  Moving on to $SU(2)_L$, $B_L$ sits in the triplet representation, so in 
terms of fundamental $SU(2)_L$ indices it is a symmetric tensor $B_{L\{\alpha\beta\}}$.  Taking the product of two $B_L$ fields, 
the $SU(2)_L$ invariant comes from contracting the indices with $\epsilon$:
\begin{align}
(B^2_L)_{(0,0)} = B_{L\{\alpha\beta\}}B_{L\{\gamma\delta\}}\epsilon^{\alpha\gamma}\epsilon^{\beta\delta}.
\label{eq:bsquared}
\end{align}
To convert from this format to the more familiar $SO(3,1)$ language (step 4), our starting point is the decomposition of 
antisymmetric tensors\,\cite{Massa:1972sc}: 
\begin{align}
X^{\mu\nu} = \sigma^{\mu\alpha\dot{\alpha}} \sigma^{\nu\beta\dot{\beta}} (\epsilon_{\dot{\alpha}\dot{\beta}}A_{\{\alpha\beta\}} 
   + \epsilon_{\alpha\beta}B_{\{\dot{\alpha}\dot{\beta}\}} ),
\label{eq:SU2SO3convert}
\end{align}
where $\mu, \nu$ are the usual $SO(3,1)$ indices and undotted (dotted) $\alpha, \beta$ are $SU(2)_L \, (SU(2)_R)$ indices.  
We can see that $A$ sits in the $(1,0)$ representation of $SU(2)_L \otimes SU(2)_R$ while $B$ sits in $(0,1)$.  Manipulating 
Eq.~\eqref{eq:SU2SO3convert} using the properties of $\eps$ and the $\sigma$ matrices\,\cite{Dreiner:2008tw}, we find
\begin{align}
A_{\{\alpha\beta\}} = 2\, i\, (\sigma^{\mu\nu})_{\alpha\beta}X_{L\mu\nu},
\quad B_{\{\dot{\alpha}\dot{\beta}\}} = 2\, i\, (\bar{\sigma}^{\mu\nu})_{\dot{\alpha}\dot{\beta}}\, X_{R\mu\nu},
\label{eq:su2toso3}
\end{align}
where $\sigma^{\mu\nu}$ and $\bar{\sigma}^{\mu\nu}$ are antisymmetric in $SO(3,1)$ indices but symmetric in either $SU(2)_L$ 
or $SU(2)_R$ indices.  Applying Eq.~\eqref{eq:su2toso3} to Eq.~\eqref{eq:bsquared} gives: 
 \begin{align}
 B_{L\{\alpha\beta\}}B_{L\{\gamma\delta\}}\epsilon^{\alpha\beta}\epsilon^{\gamma\delta} 
  &= -4\, (\sigma^{\mu\nu})_{\alpha\beta}(\sigma^{\rho\sigma})_{\gamma\delta}\,\epsilon^{\alpha\gamma}\epsilon^{\beta\delta}\, 
        \times B_{L\mu\nu}B_{L\rho\sigma}\nonumber \\
  &= -4\, {\textrm{Tr}}(\sigma^{\mu\nu}\sigma^{\rho\sigma})\, B_{L\mu\nu}B_{L\rho\sigma} \nonumber \\
  &= -2\,(\,g^{\mu\rho}g^{\nu\sigma} - g^{\mu\sigma}g^{\nu\rho} 
     -i\, \eps^{\mu\nu\rho\sigma} )\,B_{L\mu\nu}B_{L\rho\sigma} \nonumber \\
 & = -4\, (B_{L\mu\nu}B_{L}^{\mu\nu} - i\,B_{L\mu\nu}\tilde{B}_{L}^{\mu\nu}).
 \end{align}
 Next, we can convert\footnote{When converting between the $X_{L,R}$ and $X, \tilde X$ forms, the relations 
$\epsilon^{\mu\nu\rho\sigma} X_{L, \rho\sigma} = 2\,i\, X^{\mu\nu}_L,\, \epsilon^{\mu\nu\rho\sigma} X_{R, \rho\sigma} 
 = -2\, i\, X^{\mu\nu}_R$ are particularly handy.} from $B_L$ to $B$ and $\tilde B$, and $\Hdag_2$ to $\Hdag_{\bar 2}$, 
with the net result:
 \begin{align}
 \mathcal O_{H2B2} = (\Hdag H)^2\,(B_{\mu\nu}B^{\mu\nu} - i\, B_{\mu\nu}\tilde B^{\mu\nu}),
 \label{ex:H2B2}
 \end{align}
again dropping overall numerical factors and subscripts. 
 
Finally, we must impose hermiticity.  Repeating the conversion steps on the hermitian conjugate output $(\Hdag H)^2 B^2_R$, 
we get (unsurprisingly) the hermitian conjugate of Eq.~\eqref{ex:H2B2}.  Including both $\mathcal O_{H2B2}$ and 
$\mathcal O^{\dag}_{H2B2}$ into the Lagrangian, hermiticity demands that their coefficients are complex conjugates of each other, 
 \begin{align}
 \mathcal L \supset c_{H2B2}\,\mathcal O_{H2B2} + c^*_{H2B2}\, \mathcal O^{\dag}_{H2B2},
 \label{eq:H2B2form1}
 \end{align}
corresponding to two real coefficients instead of four.  We can regroup these terms into two separate operators involving Higgs 
fields and hypercharge field strengths:
 \begin{align}
 \mathcal O_1 = (\Hdag\, H)^2\, B_{\mu\nu}B^{\mu\nu},\quad  \mathcal O_2 = (\Hdag\, H)^2\, B_{\mu\nu}\tilde B^{\mu\nu}.
\label{eq:H2B2form2}
 \end{align}
Either in the form of Eq.~\eqref{eq:H2B2form1} or Eq.~\eqref{eq:H2B2form2}, we see that the result of steps $1-6$ is a number 
of operators matching the Hilbert series output.  In Eq.~\eqref{eq:H2B2form2} the two operators corresponding to the outputs 
$(\Hdag H)^2 B^2_{L,R}$ are obvious, while in Eq.~\eqref{eq:H2B2form1} they are combined into a complex, non-hermitian operator 
with a complex coefficient\footnote{Said differently, the Hilbert series output dictates the number of real operator coefficients 
for a given field content. Customarily, bosonic operators are written in the form of Eq.~\eqref{eq:H2B2form2}, with real 
coefficients, while fermionic operators are written in the form of Eq.~\eqref{eq:H2B2form1}, with complex coefficients and an 
implicit addition of the hermitian conjugate.}.  In cases where there is more than one operator with a given field content, there 
are more options for the operators (and their combinations) to use.  This can be seen with $2\, (\Hdag H)^2 W^2_L$, which also 
has non-trivial $SU(2)_w$ contractions and is presented in Appendix~\ref{app:examples}.  These choices are inevitable and 
represent a choice of basis. Note the dimension-8 CPV terms will not affect cross sections at 
$\mathcal O(1/\Lambda^4)$ since they have no SM piece to interfere with.  We include them purely for completeness and to show 
how the operator counting works at different stages of the translation process.

The steps above carry over to operators with a single derivative. The new ingredient is the derivative $D$, which transforms 
as $(\half, \half; 0,0,0)$ under the symmetry groups and is otherwise treated like any other bosonic object in an operator. 
When converting an operator of the form $D(A\, B\, C)$, where $A,B,C$ are SMEFT fields, the first conversion step is to 
partition the derivative among the fields. This is not as automatic as it seems, since i.) we are only interested in products 
of derivatives and fields that do not reduce by the equations of motion, and ii.) it is often the case that we 
cannot apply the derivative to all fields present, e.g. for $D \,ABC $ there may be no way to make an invariant with $AB\, D(C)$. 
The procedure for removing EOM-reducible terms in Hilbert series output was put forth in Ref.~\cite{Lehman:2015via}. In short, 
we treat $D\psi$ ($\psi$ a left-handed fermion) as a Lorentz $(1,\half)$ in all products. Similarly, 
$D\psi^{\dag} \sim (\half, 1)$, $DX_L \sim (\frac 3 2, \half)$ and $DX_R \sim (\half, \frac 3 2)$\footnote{There is no EOM 
reduction for scalars at ${\mathcal{O}}(D)$, where $D\phi \sim (\half, \half)$.  There is reduction at ${\mathcal{O}}(D^2)$, 
where $D^2\phi \sim (1,1)$.}.  In terms of $SU(2)_L \times SU(2)_R$ indices:
\begin{align}
D\psi \sim (D\psi)_{\{\alpha\beta\}, \dot{\alpha}}\, \quad D\psi^{\dag} \sim (D\psi^{\dag})_{\alpha, \{\dot{\alpha}\dot{\beta}\}}
\, \quad DX_L \sim (DX_L)_{\{\alpha\beta\gamma\}, \dot{\alpha}}\,\quad DX_R \sim 
(DX_R)_{\alpha,\{\dot{\alpha}\dot{\beta}\dot{\gamma}\}} \nonumber
\end{align}
and we can convert between $SU(2)_L \times SU(2)_R$ and $SO(3,1)$ using
\begin{align}
D_{\alpha\dot{\alpha}} = D_{\mu} (\sigma^{\mu})_{\alpha\dot{\alpha}}.
\end{align}
In the manipulations above we have only shown the Lorentz part of the representations.  Once partitioned onto a field, the 
derivative should be thought of as a covariant derivative, so $D\psi,~DX_L$, etc. will carry the gauge representations 
appropriate to $\psi,~X_L$, etc.
 
Once we know which derivative partitions are allowed, we are free to pick which one to use since they are easily related to 
each other by integration by parts.  For example, if $D(ABC) = D(A) B C + A\, D(B) C$ then we may pick whichever we like 
as our operator. Once we have chosen how to partition the derivative, we proceed with steps $1-6$ of Sec.~\ref{sec:D0} to 
convert them into operators. There are no purely bosonic operators in the SMEFT containing a single derivative, therefore 
we defer an example of an $\mathcal O(D)$ operator (involving fermions) to Appendix~\ref{app:examples}.

\subsection{Operators with two or more derivatives}

Operators containing two or more derivatives are trickier. The presence of derivatives implies that we have redundancies due 
to integration by parts, which can shift the covariant derivative from one field to another, and the equations of motion.  We 
must be careful to ensure that our final result does not contain any redundancies, but has enough flexibility to generate 
(via IBP) any operator with the same fields and number of derivatives.  As before, these issues are best demonstrated with an 
example.

\subsubsection{Example: $2\, D^2(H^\dagger H B_L W_L)$}

Let us look at one of the classes of operators from Eq.~\eqref{eq:HS2}: $2\,D^2(H^\dagger H B_L W_L)$. The Hilbert series 
tells us that there are only two independent invariants for this combination of fields and derivatives, but which two operators 
do we pick? At first glance there are multiple ways of placing the derivatives, 
\begin{equation}
\begin{gathered}
\label{eq:exampleD2H2BW}
(D_\mu H^\dagger) (D_\mu H) W_{L,\rho\sigma}  B_{L,\rho\sigma}, \enskip (D_\mu H^\dagger)(D^\nu H) W_{L,\mu\rho} B_{L,\rho\nu}, \\
H^\dagger H (D_\mu W_{L,\nu\rho}) (D^\mu B_L^{\nu\rho}), \enskip (D_\mu H^\dagger)H(D_\mu W_{L,\nu\rho})B_{L,\nu\rho}, \;\; \dots
\end{gathered}
\end{equation}

If all operators were equal, we could just pick any two. However, this is not the case.  After picking one operator from 
Eq.~\eqref{eq:exampleD2H2BW} there are some choices for the second operator which---combined with the original operator---can 
transform via IBP into any of the other $D^2(\Hdag H B_L W_L)$ operators, while there are other operators that will only 
transform into a subset.  Said another way, we need to pick two operators that {\em any} of the possible $D^2(\Hdag H B_L W_L)$ 
operators can be reduced to by successive IBP. In order to make the right choice in this example and in similar cases, we need 
to know how IBP relates all operators with a given field and derivative content.

To systematically understand the IBP relations and their use in reducing the number of operators, 
we will follow the approach described in Ref.~\cite{Lehman:2015coa}. Our first step is to enumerate the ways to 
partition the derivatives, as partially illustrated in Eq.~\eqref{eq:exampleD2H2BW}.  Given our previous experience with adding 
indices, we can immediately recognize that i.) $D^2\Hdag$ and $D^2H$ will never admit a Lorentz singlet since all other fields 
only transform under Lorentz $SU(2)_L$, and ii.) the $SU(2)_w$ part of the index contraction is trivial, as $\Hdag_i H_j$  must 
form a triplet to contract with $W_L$. Partitioning the derivatives all possible ways, there are seven different operators in 
$D^2(\Hdag H B_L W_L)$. The operators are listed in Table~\ref{table:D2} with $SU(2)_w$ indices suppressed.

\begin{table}[t!]
\centering
\begin{tabular}{c|c}
$x_1$ & $(D\Hdag)_{\alpha\dot{\alpha}}(DH)_{\beta\dot{\beta}}B_{L\{\gamma\delta\}}W_{L\{\xi \eta\}}\,\epsilon^{\dot{\alpha}\dot{\beta}}\epsilon^{\alpha\beta}\epsilon^{\gamma \xi}\epsilon^{\delta \eta}$ \\
$x_2$ & $(D\Hdag)_{\alpha\dot{\alpha}}(DH)_{\beta\dot{\beta}}B_{L\{\gamma\delta\}}W_{L\{\xi \eta\}}\,\frac 1 2 \epsilon^{\dot{\alpha}\dot{\beta}}\epsilon^{\delta \xi}(\epsilon^{\alpha\gamma}\epsilon^{\beta \eta} + \epsilon^{\beta\gamma}\epsilon^{\alpha \eta}$) \\
$x_3$ & $(D\Hdag)_{\alpha\dot{\alpha}}\, H\, (DB_L)_{\{\beta\gamma\delta\},\dot{\beta}} W_{L\{\xi \eta\}}\, \epsilon^{\dot{\alpha}\dot{\beta}}\epsilon^{\alpha\beta}\epsilon^{\gamma \xi}\epsilon^{\delta \eta}$ \\
$x_4$ & $(D\Hdag)_{\alpha\dot{\alpha}}\, H\, B_{L\{\xi \eta\}} (DW_L)_{\{\beta\gamma\delta\},\dot{\beta}} \, \epsilon^{\dot{\alpha}\dot{\beta}}\epsilon^{\alpha\beta}\epsilon^{\gamma \xi}\epsilon^{\delta \eta}$ \\
$x_5$ & $\Hdag\, (DH)_{\alpha\dot{\alpha}}\, (DB_L)_{\{\beta\gamma\delta\},\dot{\beta}} W_{L\{\xi \eta\}}\, \epsilon^{\dot{\alpha}\dot{\beta}}\epsilon^{\alpha\beta}\epsilon^{\gamma \xi}\epsilon^{\delta \eta}$ \\
$x_6$ & $\Hdag\, (DH)_{\alpha\dot{\alpha}}\, B_{L\{\xi \eta\}} (DW_L)_{\{\beta\gamma\delta\},\dot{\beta}} \, \epsilon^{\dot{\alpha}\dot{\beta}}\epsilon^{\alpha\beta}\epsilon^{\gamma \xi}\epsilon^{\delta \eta}$ \\
$x_7$ & $\Hdag H\, (DB_L)_{\{\alpha\beta\gamma\},\dot{\alpha}}(DW_L)_{\{\xi \eta \delta \},\dot{\beta}}\, \epsilon^{\dot{\alpha}\dot{\beta}}\epsilon^{\alpha \xi}\epsilon^{\beta \eta}\epsilon^{\gamma \delta}$
\end{tabular}
\caption{Operators of the type $D^2(\Hdag H B_L W_L)$ where we have ignored IBP relations between terms. We have neglected all 
$SU(2)_w$ indices since there is only one possible contraction.}
\label{table:D2}
\end{table}
The first two operators correspond to the two ways we can pair $(D\Hdag\,DH) \supset (0 \oplus 1, 0 \oplus 1; 0, 0 \oplus 1,0)$ 
with $B_L\, W_L \supset (0 \oplus 1 \oplus 2, 0; 0,1,0)$\footnote{In the first operator the Lorentz indices of $D\Hdag DH$ are 
stitched together to form a singlet, as are the indices of $B_L\, W_L$.  In the second operator, we pick out the $(1,0)$ part of 
$B_L W_L$ by contracting one index on each field strength together, then combine that object with the $(1,0)$ piece of 
$D\Hdag DH$. The two operators represent the two different ways to tie the indices of $D\Hdag DH$ to the indices of $B_L W_L$; we 
could collapse the two operators to one (plus a piece looking like $x_1$) via the Schouten identity, but the current form makes 
the algebra easier.}.  We will stick to the $SU(2)_L \otimes SU(2)_R$ form of Lorentz symmetry throughout to avoid translating 
derivatives of field strengths into $SO(3,1)$ language. When contracting indices we have made a choice of the overall sign. 
Nothing will depend on this choice, but we do need to be careful to stick with this convention.

Next, we need the set of operators with one less derivative, $D(\Hdag H B_L W_L)$ that sits in the four-vector Lorentz 
representation. The group theory here follows exactly as before, except that we are picking products in the 
$(\frac 1 2, \frac 1 2)$ representation rather than Lorentz singlets. The single derivative can act on each of the four 
fields, $D\Hdag\cdots, \Hdag DH\cdots, \Hdag H (DB)$, etc., and for the $D\Hdag, DH$ options there are two ways to form 
$(\frac 1 2, \frac 1 2)$.  Working this out generates the six $D(\Hdag H B_L W_L)$ terms in Table~\ref{table:D1}.
\begin{table}[h!]
\centering
\begin{tabular}{c|c}
$y_1$ & $(D\Hdag)_{\alpha\dot{\alpha}}\,H\, B_{L\{\beta\gamma\}}W_{L\{\xi \eta\}}\, \epsilon^{\beta \xi}\epsilon^{\gamma \eta}$ \\
$y_2$ & $(D\Hdag)_{\alpha\dot{\alpha}}\,H\, B_{L\{\beta\gamma\}}W_{L\{\xi \eta\}}\, \frac 1 2 \epsilon^{\gamma \xi}(\epsilon^{\alpha\beta} + \epsilon^{\alpha \eta})$ \\
$y_3$ & $\Hdag\, (DH)_{\alpha\dot{\alpha}}\, B_{L\{\beta\gamma\}}W_{L\{\xi \eta\}}\, \epsilon^{\beta \xi}\epsilon^{\gamma \eta}$ \\
$y_4$ & $\Hdag\, (DH)_{\alpha\dot{\alpha}}\, B_{L\{\beta\gamma\}}W_{L\{\xi \eta\}}\, \frac 1 2 \epsilon^{\gamma \xi}(\epsilon^{\alpha\beta} + \epsilon^{\alpha \eta})$ \\
$y_5$ & $\Hdag\, H\, (DB_L)_{\{\alpha\beta\gamma\},\dot{\alpha}}\, W_{L\{\xi \eta\}}\, \epsilon^{\beta \xi}\epsilon^{\gamma \eta} $ \\
$y_6$ & $\Hdag\, H\,  B_{L\{\xi \eta\}} (DW_L)_{\{\alpha\beta\gamma\},\dot{\alpha}}\, \epsilon^{\beta \xi}\epsilon^{\gamma \eta} $
\end{tabular}
\caption{Operators of the type $D(\Hdag H B_L W_L)$ that are gauge-invariant but sit in the Lorentz four-vector representation. 
The number of operators in this class can be generated automatically via the same procedure that projects out the number of 
total invariants.  As in Table~\ref{table:D2}, we have suppressed $SU(2)_w$ indices.}
\label{table:D1}
\end{table}

The relation between the $y_i$---gauge invariant operators with one fewer derivative and sitting in the four-vector Lorentz 
representation---and IBP is now easy to see. Contracting any of the $y_i$ with a final derivative results in a linear 
combination of $D^2(\Hdag H B_L W_L)$ operators making up a total derivative.  Therefore, each $D(y_i)$ equation provides an 
IBP relation among the higher-derivative terms. For example:
\begin{align}
D_{\delta\dot{\delta}} (y_1) &= \text{total deriv.}  = D_{\delta\dot{\delta}}\Big( (D\Hdag)_{\alpha\dot{\alpha}}\,H\, B_{L\{\beta\gamma\}}W_{L\{\xi \eta\}}\, \epsilon^{\beta x}\epsilon^{\gamma y}  \Big) \epsilon^{\dot{\alpha}\dot{\delta}}\epsilon^{\alpha \delta}\ \nonumber \\
& = \Big( (D\Hdag)_{\alpha\dot{\alpha}}\, (DH)_{\delta\dot{\delta}}\, B_{\{\beta\gamma\}}W_{\{\xi \eta\}} + (D\Hdag)_{\alpha\dot{\alpha}}\, H\, (DB_L)_{\{\delta \beta\gamma\},\dot{\delta}}\,W_{L\{\xi \eta\}} + \nonumber \\
& \quad\quad\quad (D\Hdag)_{\alpha\dot{\alpha}}\, H\, B_{L\{\beta\gamma\}}\, (DW_L)_{\{\delta \xi \eta\},\dot{\delta}} \Big)\,  \epsilon^{\dot{\alpha}\dot{\delta}}\epsilon^{\alpha \delta}\epsilon^{\beta \xi}\epsilon^{\gamma \eta} \nonumber \\
& = x_1 + x_3 + x_4,
\label{eq:IBP}
\end{align}
where we will usually have to do some index juggling to get the contractions in $D(y_1)$ to match those of the $x_i$. Notice 
that in addition to adding the derivative, we have to specify (and stick to) a convention on how to stitch up the remaining 
indices. The net result of $D(y_1)$ is that $x_1, x_3$ and $x_4$ are not all independent, i.e. given two we can generate the 
third. 

Following this logic, each of the other $D(y_i)$ provides a relation, or constraint, among the $x_i$.  If each of the $D(y_i)$ 
were independent, this would tell us that the true number of independent operators---including IBP relations---is $\# D^2$ 
operators$- \# D$ operators, $\# x_i - \# y_i$, or more generally $(\#$ operators at $\mathcal O(D^m)) - (\#$ operators at 
$\mathcal O(D^{m-1})$).  However, in practice, the constraint equations are often redundant, making the number of independent 
constraints $< \# y_i$. 

To get at the number of independent constraints, we can write the constraint equations as a matrix with each $D(y_i)$ as a 
row acting on a vector of $x_i$, and then determine its rank. For the example here, carrying out the same manipulations as 
in Eq.~\eqref{eq:IBP} for the rest of the $y_i$, we find:
\begin{align}
\left(\begin{array}{ccccccc}
1 & 0 & 1& 1 & 0 & 0 & 0 \\
0 & -1 & -\frac 1 2 & \frac 1 2 & 0 & 0 & 0 \\
1& 0 & 0 & 0 & 1 & 1 & 0 \\
0 & 1 & 0 & 0 & -\frac 1 2 & \frac 1 2 & 0 \\
0 & 0 & 1 & 0 & 1 & 0 & 1 \\ 
0 & 0 & 0 & 1 & 0 & 1& 1\\
\end{array}\right)\left(\begin{array}{c} x_1 \\ x_2 \\ x_3 \\ x_4 \\ x_5 \\ x_6 \\ x_7 \end{array}\right) 
  = M_{IBP}\cdot \vec{x} = 0
\end{align}

The constraint matrix $M_{IBP}$ has rank 5, indicating only five of the six IBP relations are actually independent. Applying five 
constraints to seven operators leaves us with two independent operators, in agreement with the Hilbert series counting.  While 
tedious, this constraint procedure can be applied to any operator type (fermion or bosons, $D>1$), including those with multiple 
electroweak contractions.  After this treatment of IBP relations was put forward in\,\cite{Lehman:2015coa} and applied to the 
SMEFT, later work\,\cite{Henning:2015alf} showed it had missed some operators by overcounting IBP relations.  Revisiting the 
constraint procedure here, we find that the error in Ref.\,\cite{Lehman:2015coa} did not lie in the method, but was due to 
mathematical mistakes made when applying the method\footnote{In particular, some of the error in Ref.~\cite{Lehman:2015coa} 
can be traced to a faulty shortcut the authors used to determine when a full matrix/rank treatment of the constraints was 
necessary, while in other circumstances it was just algebraic error. Clearly this method would benefit from automation, 
possibly along the lines of Ref.~\cite{Gripaios:2018zrz}.}.  When applied correctly, the number of independent operators 
(after IBP) found using the constraint method agrees with the Hilbert series counting (at least at dimension $\le 8$). The 
payoff of the constraint method is that it gives us the actual form of the operators and tells us which operators are related 
by IBP and which are not.  For the example at hand, after row-reducing $M_{IBP}$ we find it can be distilled to the following 
relations:
\begin{align}
x_1 - x_7 = 0,\, x_2 + x_6 + \frac{x_7}{2} = 0,\, x_3 - x_6 = 0, \nonumber \\
x_4 + x_6 + x_7 = 0,\, x_5 + x_6 + x_7 = 0. \nonumber 
\end{align}
From these relations, we see that $x_1$ and $x_2$ are sufficient to generate all seven operators.  Thus, given a combination 
$c_i x_i$, we can IBP repeatedly (throwing away surface terms) and collapse the sum into $c_{1,eff}\, x_1 + c_{2,eff}\, x_2$, 
with $c_{eff}$ some linear combination of the initial $c_i$.  We could collapse the sum into other pairs of operators, such as 
$\{x_1, x_3\}$ or $\{x_6, x_7\}$, however there are also other pairs, such as $\{ x_4, x_5\}$, that we could not 
reduce to.  Choosing $\{x_1, x_2\}$ to span the set, reintroducing the $SU(2)_w$ indices, and performing steps 4 and 6 from 
Sec.~\ref{sec:D0}, we are left with:
\begin{align}
\mathcal O_{2DH2BW1} &= \Tr(D_{\mu}\Hdag\,\tau^I\,D^{\mu}H)(B_{L}^{\rho\sigma}W^I_{L,\rho\sigma} - i B_{L}^{\rho\sigma}\tilde W^I_{L,\rho\sigma}) \nonumber\\ 
\mathcal O_{2DH2BW2} &= \Tr(D_{\mu}\Hdag\,\tau^I\,D_{\nu}H)(B_{L}^{\mu\kappa}W^{I\,\nu}_{L,\kappa} - B_{L}^{\nu\kappa}W^{I\,\mu}_{L,\kappa} + \frac i 2 B^{\nu\kappa}_{L}\tilde W^{\mu}_{L,\kappa} - \frac i 2 B^{\mu\kappa}_L\tilde W^{\nu}_{L,\kappa} \nonumber \\
& \quad\quad\quad \quad\quad\quad\quad\quad \quad\quad\quad \quad \quad\quad\quad\quad\quad\quad + \frac i 2 \tilde B^{\nu\kappa}_L\, W^{\mu}_{L,\kappa} - \frac i 2 \tilde B^{\mu\kappa}_L\, W^{\nu}_{L,\kappa} ),
\end{align}
where, as in Eq.~\eqref{ex:H2B2}, we have converted to $\Hdag_{\bar 2}$ format.

One remaining question is the origin of the IBP relation redundancies. Combining several of the $y_i$ and manipulating indices, 
we see that some combinations of the constraint operators can be expressed as a total derivative:
\begin{align}
y_2 + y_4 + \frac 1 2 (y_5 - y_6) = D_{\alpha\dot{\alpha}}\Big(\Hdag\, H\,B_{\{\beta\gamma\}}W_{\{\xi \eta\}}\, 
\frac 1 2 \epsilon^{\gamma \xi}(\epsilon^{\alpha\beta} + \epsilon^{\alpha \eta})\Big)
\label{eq:IBPrel}
\end{align}
When we apply a final derivative, Eq.~\eqref{eq:IBPrel} connects the constraints from $D(y_2)$, $D(y_4)$, $D(y_5)$, and $D(y_6)$ 
so they are no longer independent. For the case here, we have a single relation among the $y_i$, so the number of independent 
constraints is reduced by one, from six to five.  To better understand why this occurs, notice that the operator in the 
parentheses of Eq.~\eqref{eq:IBPrel} transforms as a $(0,1)$ Lorentz representation.  Applying $D^2$ to this combination will 
always give zero, since $D^2$ has Lorentz irreducible representations $(0,0) \oplus (1,1)$, which cannot form a singlet with 
$(0,1)$.  Enforcing this fact---that $D^2 \otimes (0,1) = 0$---results in relations among the $\mathcal O(D)$ 
operators\footnote{Technically, we are only interested in $(0,1)$ or $(1,0)$ operators that are not themselves a total 
derivative. The Hilbert series iteratively removes total derivative terms, as explained in~\cite{Henning:2015alf}. In the 
constraint method shown here, $D^{m-2}, D^{m-3}\,\cdots$ total derivatives all show up as relations among rows of $M_{IBP}$.}.

\section{Dimension-8 operators relevant for $pp \to h\, W$}
\label{sec:bos8}

Having worked through a few examples, we now present the results of the operator extraction for the full set of terms listed 
in Eqs.~$\eqref{eq:HS1}-\eqref{eq:HS4}$. 

\begin{itemize}
\item For the bosonic terms there are a total of 17 operators with no derivatives, corresponding to the eight terms in 
Eq.~\eqref{eq:HS1} and their hermitian conjugates, along with the single self-hermitian term.  The set is listed in 
Table~\ref{17_derivative_free}.  As explained in Sec.~\ref{sec:D0}, we have made some choices about how to display indices 
(e.g. triplets vs. doublets) and what linear combinations to take to form operators with simple properties under CP 
transformations.  These choices constitute a choice of basis.

\item Table~\ref{26_two_derivative} contains the bosonic operators with two derivatives.  There are 26 operators, corresponding 
to ten terms in Eq.~\eqref{eq:HS2} plus their hermitian conjugates, and the six self-hermitian terms.  Because there are 
derivatives, there is even more choice than in Table~\ref{17_derivative_free}.  When possible, we have opted to put the 
derivatives on the Higgs fields as this makes implementing the operators into FeynRules easier.

\begin{table}[t!]
\begin{center}
\begin{tabular}{|c|c|c|c|} \hline
 $ \mathcal{O}_{8,H} $ & $ (H^\dagger H)^4 $ 
  & $ \mathcal{O}_{8,W} $ & $ \epsilon_{IJK}\, (H^\dagger H) W^{\mu\nu,I} W_{\nu\rho}^J W_{\mu}^{\rho,K} $ \\
 $ \mathcal{O}_{8,HB} $ & $ (H^\dagger H)^2 B_{\mu\nu}B^{\mu\nu} $  
  & $ \mathcal{O}_{8,\tilde{W}} $ & $ \epsilon_{IJK}\, (H^\dagger H) W^{\mu\nu,I} \Wtilde_{\nu\rho}^J W_{\mu}^{\rho,K} $ \\
 $ \mathcal{O}_{8,H\tilde{B}} $ & $ (H^\dagger H)^2 B_{\mu\nu} \Btilde^{\mu\nu} $ 
  &  $\mathcal{O}_{8,HG§} $ & $ \delta_{AB}\, (H^\dagger H)^2 G_{\mu\nu}^A G^{\mu\nu, B}$ \\
  $ \mathcal{O}_{8,HWB} $ & $ \delta_{IJ}\, (H^\dagger H)(H^\dagger \tau^I H) B_{\mu\nu}W^{\mu\nu, J} $ 
  & $ \mathcal{O}_{8,H\tilde{G}} $ & $ \delta_{AB}\, (H^\dagger H)^2 G_{\mu\nu}^A\Gtilde^{\mu\nu,B} $\\
 $ \mathcal{O}_{8,H\tilde{W}B} $ & $ \delta_{IJ}\, (H^\dagger H)(H^\dagger \tau^I H) B_{\mu\nu}\Wtilde^{\mu\nu, J} $ 
  &  $ \mathcal{O}_{8,G} $ & $ f_{ABC}\, (H^\dagger H) G^{\mu\nu,A} G_{\nu\rho}^B G_{\mu}^{\rho,C} $ \\
  $ \mathcal{O}_{8,HW} $ & $ \delta_{IJ} (H^\dagger H)^2 W_{\mu\nu}^I W^{\mu\nu, J}$ 
  &  $ \mathcal{O}_{8,\tilde{G}} $ & $ f_{ABC}\, (H^\dagger H) G^{\mu\nu,A} \Gtilde_{\nu\rho}^B G_{\mu}^{\rho,C} $ \\
  \cline{3-4}
 $ \mathcal{O}_{8,H\tilde{W}} $ & $ \delta_{IJ} (H^\dagger H)^2 W_{\mu\nu}^I\Wtilde^{\mu\nu, J} $ \\
 $ \mathcal{O}_{8,HW2} $ & $ \delta_{IK} \delta_{JM} (H^\dagger \tau^I H) (H^\dagger\tau^J H) W_{\mu\nu}^K W^{\mu\nu, M} $\\
 $  \mathcal{O}_{8,H\tilde{W}2} $ & $ \delta_{IK} \delta_{JM} (H^\dagger \tau^I H) (H^\dagger\tau^J H) W_{\mu\nu}^K \Wtilde^{\mu\nu, M} $ \\
 $ \mathcal{O}_{8,HWB2} $ & $ \epsilon_{IJK}\, (H^\dagger \tau^I H) B_{\mu}^{\nu} W_{\nu\rho}^J W^{\mu\rho,K} $ \\
 $ \mathcal{O}_{8,HW\tilde{B}2} $ & $ \epsilon_{IJK}\, (H^\dagger \tau^I H) \left(\Btilde^{\mu\nu}W_{\nu\rho}^J 
  W_{\mu}^{\rho,K} + B^{\mu\nu} W_{\nu\rho}^J \Wtilde_{\mu}^{\rho,K} \right) $ \\
 \cline{1-2}
\end{tabular}
\caption{The 17 derivative-free operators after conversion to the standard $X, \widetilde{X}$ notation for the field-strength 
tensors and with $\Hdag$ in the $\bar 2$ representation.}
\label{17_derivative_free}
\end{center}
\end{table}

\begin{table}[t!]
\begin{center}
\footnotesize
\begin{tabular}{|c|c|c|c|}
\hline
$ \mathcal{O}_{8,HD} $ & $(H^\dag H)^2(D_\mu H^\dag \, D^\mu H)$  
 & $\mathcal{O}_{8,DH\tilde{W}3b} $ & $\epsilon_{IJK}\, (D^\mu H^\dag \tau^I D^\nu H)(W^J_{\mu\rho}\Wtilde^{\rho,K}_{\nu} + \Wtilde^J_{\mu\rho} W^{\rho,K}_{\nu})  $ \\
$ \mathcal{O}_{8,HD2} $ & $\delta_{IJ}\, (H^\dag H)(H^\dag \tau^I H)(D^\mu H^\dag \tau^J D_\mu H) $ 
 & $\mathcal{O}_{8,DHWB} $ & $\delta_{IJ}\, (D^\mu H^\dag\, \tau^I D_\mu H)
B^{\rho\sigma}W^J_{\rho\sigma}$ \\
$ \mathcal{O}_{8,DHB} $ & $(D^{\mu} H^\dag \,D^{\nu}H)B_{\mu\rho}B^{\rho}_{\nu} $ 
 & $\mathcal{O}_{8,DH\tilde{W}B} $ & $\delta_{IJ}\, (D^\mu H^\dag\, \tau^I D_\mu H) B^{\rho\sigma} \Wtilde^J_{\rho\sigma}$ \\
$ \mathcal{O}_{8,DHB2} $ & $ (D^\mu H^\dag D_\mu H)B^{\rho\sigma}B_{\rho\sigma}$ 
 & $\mathcal{O}_{8,DHWB2} $ & $i\,\delta_{IJ}\, (D^\mu H^\dag \tau^I D^\nu H)(B_{\mu\rho}W^{\rho,J}_{\nu} - B_{\nu\rho}W^{\rho,J}_{\mu})  $ \\
$ \mathcal{O}_{8,DH\tilde{B}2} $ & $ (D^\mu H^\dag D_\mu H) B^{\rho\sigma}\Btilde_{\rho\sigma}$ 
 & $\mathcal{O}_{8,DHWB3} $ & $\delta_{IJ}\, (D^\mu H^\dag \tau^I D^\nu H)(B_{\mu\rho}W^{\rho,J}_{\nu} + B_{\nu\rho}W^{\rho,J}_{\mu}) $ \\
$ \mathcal{O}_{8,DHG} $ & $ \delta_{AB}\, (D^\mu H^\dag D^\nu H) G^A_{\mu\rho}G^{\rho,B}_{\nu}$ 
 & $\mathcal{O}_{8,DH\tilde{W}B2} $ & $\delta_{IJ}\, (D^\mu H^\dag \tau^I D^\nu H)
(B^{\rho}_{[\mu}\Wtilde^J_{\nu]\rho} - \Btilde^{\rho}_{[\mu}W^J_{\nu]\rho}) $\\
$ \mathcal{O}_{8,DHG2} $ & $ \delta_{AB}\, (D^\mu H^\dag D_\mu H) G^{\rho\sigma,A}G^B_{\rho\sigma}$  
 & $\mathcal{O}_{8,DH\tilde{W}B3} $ & $\delta_{IJ}\, (D^\mu H^\dag \tau^I D^\nu H)
(B^{\rho}_{\{\mu}\Wtilde^J_{\nu\}\rho} + \Btilde^{\rho}_{\{\mu}W^J_{\nu\}\rho})  $ \\
$ \mathcal{O}_{8,DH\tilde{G}2} $ & $\delta_{AB}\, (D^\mu H^\dag D_\mu H) G^{\rho\sigma,A}\Gtilde^B_{\rho\sigma}$ 
 & $\mathcal{O}_{8,HDHB} $ & $i\,(H^\dag H)(D_\mu H^\dag\, D_\nu H) B^{\mu\nu}$ \\
$ \mathcal{O}_{8,DHW} $ & $\delta_{IJ}\, (D^\mu H^\dag D^\nu H)
W^I_{\mu\rho}W^{\rho,J}_{\nu}$
 & $\mathcal{O}_{8,HDH\tilde{B}} $ & $i\,(H^\dag H)(D_\mu H^\dag\, D_\nu H) \Btilde^{\mu\nu}$ \\
$ \mathcal{O}_{8,DHW2}$ & $\delta_{IJ}\, (D^\mu H^\dag D_\mu H) W^{\rho\sigma,I} W^J_{\rho\sigma}$ 
 & $\mathcal{O}_{8,HDHW} $ & $i\, \delta_{IJ}\, (H^\dag H) (D^\mu H^\dag \tau^I D^\nu H)W^J_{\mu\nu}$ \\
$ \mathcal{O}_{8,DH\tilde{W}2}$ & $\delta_{IJ}\, (D^\mu H^\dag D_\mu H)W^{\rho\sigma,I}\Wtilde^J_{\rho\sigma}$ 
 & $\mathcal{O}_{8,HDH\tilde{W}}$ & $i\,\delta_{IJ}\, (H^\dag H) (D^\mu H^\dag \tau^I D^\nu H) \Wtilde^J_{\mu\nu}  $ \\
$ \mathcal{O}_{8,DHW3}$ & $\epsilon_{IJK}\, (D^\mu H^\dag \tau^I D^\nu H) W^J_{\mu\rho}W^{\rho,K}_{\nu} $ 
 & $\mathcal{O}_{8,HDHW2} $ & $i\,\epsilon_{IJK}\, (H^\dag \tau^I H) (D^\mu H^\dag \tau^J D^\nu H)W^K_{\mu\nu}$ \\
$ \mathcal{O}_{8,DH\tilde{W}3a}$ & $ \epsilon_{IJK}\, (D^\mu H^\dag \tau^I D^\nu H) 
(W^J_{\mu\rho}\Wtilde^{\rho,K}_{\nu} - \Wtilde^J_{\mu\rho}W^{\rho,K}_{\nu})  $  
 & $\mathcal{O}_{8,HDH\tilde{W}2} $ & $i\,\epsilon_{IJK}\, (H^\dag \tau^I H) (D^\mu H^\dag \tau^J D^\nu H) \Wtilde^K_{\mu\nu}$ \\
\hline
\end{tabular}
\caption{The 26 two-derivative operators after conversion to the standard $X$,
	$\widetilde{X}$ notation (plus linear combinations). Factors of $i$ are
	included where necessary so that the operators are explicitly
	self-hermitian with real coefficients. 
		}
\label{26_two_derivative}
\end{center}
\end{table}

\item Table~\ref{four_derivs} contains the three bosonic operators at $\mathcal O(D^4)$.  When forming these operators 
(and all others in this section) we have ignored overall signs or numerical factors.
\begin{table}[h!]
\begin{center}
\begin{tabular}{|c|c|} 
\hline
$\mathcal O_{8,4D1}$ & $(D_\mu H^\dagger D_\nu H)(D^\nu H^\dagger D^\mu H)$ \\
$\mathcal O_{8,4D2}$ & $(D_\mu H^\dagger D_\nu H)(D^\mu H^\dagger D^\nu H)$  \\
$\mathcal O_{8,4D3}$ & $(D^\mu H^\dagger D_\mu H)(D^\nu H^\dagger D_\nu H)$ \\ \hline
\end{tabular}
\caption{The explicit forms of the three bosonic dimension-8 operators containing at least one Higgs field and 
four derivatives. These three operators have three independent real coefficients.}
\label{four_derivs}
\end{center}
\end{table}

\begin{table}[h!]
\begin{center}
\begin{tabular}{|c|c|c|c|}
\hline
$\mathcal O_{8,QW1}$ & $\delta_{IJ}\,(Q^{\dag}\bar{\sigma}^{\nu} Q)\,D^{\mu}(\Hdag \tau^I H)\, W^J_{\mu\nu}$ 
  & $\mathcal O_{8,Q1}$ & $i (Q^{\dag}\bar{\sigma}^{\mu} Q)(\Hdag\,\overleftrightarrow D^{\mu} H)(\Hdag H)$\\
$\mathcal O_{8,Q\tilde W1}$ & $ \delta_{IJ}\,(Q^{\dag}\bar{\sigma}^{\nu} Q)\,D^{\mu}(\Hdag \tau^I H)\, \tilde W^J_{\mu\nu}$ 
  & $\mathcal O_{8,Q2}$ & $i\,\delta_{IJ}\,(Q^{\dag}\bar{\sigma}^{\mu} \tau^I\, Q)\Big((\overleftrightarrow D_{\mu}\Hdag \tau^J H) (\Hdag H) +$ \\
& & & $\quad\quad (\overleftrightarrow D_{\mu}\Hdag\, H)(\Hdag \tau^J\, H)\Big)$ \\
 $\mathcal O_{8,QW2}$ & $i\,\delta_{IJ}\,(Q^{\dag}\bar{\sigma}^{\nu} Q)(\Hdag\overleftrightarrow D^{\mu} \tau^I H)\, W^J_{\mu\nu}$ 
  & $\mathcal O_{8,Q3}$ & $i\,\epsilon_{IJK}(Q^{\dag}\bar{\sigma}^{\mu} \tau^I\, Q)(\Hdag \overleftrightarrow D^{\mu} \tau^J H)(\Hdag \tau^K\, H)$\\
 $\mathcal O_{8,Q\tilde W2}$ & $i\,\delta_{IJ}\,(Q^{\dag}\bar{\sigma}^{\nu} Q)(\Hdag \overleftrightarrow D^{\mu} \tau^I H)\, \tilde W^J_{\mu\nu}$ 
  & $\mathcal O_{8,Q4}$ & $\epsilon_{IJK}(Q^{\dag}\bar{\sigma}^{\mu} \tau^I\, Q)(\Hdag \tau^J H)\, D_{\mu}(\Hdag \tau^K\, H)$  \\
 \cline{3-4}
$\mathcal O_{8,QW3}$ & $\delta_{IJ}\,(Q^{\dag}\bar{\sigma}^{\nu}\tau^I\,Q)\,D^{\mu}(\Hdag H)\, W^J_{\mu\nu}$ 
  & $\mathcal O_{8,3Q1}$ & $i\,(Q^{\dag} \bar{\sigma}^\mu D^\nu Q)(D^2_{(\mu\nu)} H^\dagger H) + h.c.$ \\
$\mathcal O_{8,Q\tilde W 3}$ & $\delta_{IJ}\,(Q^{\dag}\bar{\sigma}^{\nu}\tau^I\,Q)\,D^{\mu}(\Hdag H)\, \tilde W^J_{\mu\nu}$ 
  & $\mathcal O_{8,3Q2}$ & $i\,\delta_{IJ}(Q^{\dag} \bar{\sigma}^\mu \tau^I D^\nu Q)(D^2_{(\mu\nu)} H^\dagger \tau^J H) + h.c.$ \\
$\mathcal O_{8,QW4}$ & $i\,\delta_{IJ}\, (Q^{\dag}\bar{\sigma}^{\nu}\tau^I\, Q)(\Hdag \overleftrightarrow D^{\mu} H)\, W^J_{\mu\nu}$ 
  & $\mathcal O_{8,3Q3}$ & $i\,(Q^{\dag} \bar{\sigma}^\mu D^\nu Q)(H^\dagger D^2_{(\mu\nu)} H) + h.c.$ \\
$\mathcal O_{8,Q\tilde W 4}$ & $i\,\delta_{IJ}\,(Q^{\dag}\bar{\sigma}^{\nu}\tau^I\, Q)(\Hdag \overleftrightarrow D^{\mu} H)\, \tilde W^J_{\mu\nu}$ 
  & $\mathcal O_{8,3Q4}$ & $i\,\delta_{IJ}(Q^{\dag} \bar{\sigma}^\mu \tau^I D^\nu Q)( H^\dagger \tau^J D^2_{(\mu\nu)} H) + h.c.$ \\
\cline{3-4}
$\mathcal O_{8,QW5}$ & $\epsilon_{ABC}\, (Q^{\dag}\bar{\sigma}^{\nu}\tau^A\ Q)\,D^{\mu}(\Hdag \tau^B H)\, W^C_{\mu\nu}$\\
$\mathcal O_{8,Q\tilde W5}$ & $\epsilon_{ABC}\, (Q^{\dag}\bar{\sigma}^{\nu}\tau^A\,Q)\, D^{\mu}(\Hdag \tau^B H)\, \tilde W^C_{\mu\nu}$ \\
$\mathcal O_{8,QW6}$ & $i\, \epsilon_{ABC}\, (Q^{\dag}\bar{\sigma}^{\nu}\tau^A Q)(\Hdag \overleftrightarrow D^{\mu} \tau^B H)\, W^C_{\mu\nu}$\\
$\mathcal O_{8,Q\tilde W6}$ & $i\, \epsilon_{ABC}\, (Q^{\dag}\bar{\sigma}^{\nu}\tau^A Q)(\Hdag \overleftrightarrow D^{\mu} \tau^B H)\, \tilde W^C_{\mu\nu}$ \\
\cline{1-2}
\end{tabular}
\end{center}
\caption{The 20 operators involving quark, $W$-boson and Higgs fields that are relevant for the phenomenological study 
of $pp \rightarrow h\,W^\pm$. }
\label{contact_operators}
\end{table}

\item Finally, Table~\ref{contact_operators} contains the fermionic operators that contribute to $pp \to h\, W$, either 
by contributing to the $\bar q q W$ vertices or through direct four-point contact terms. There are no terms with even 
numbers of derivatives, as operators of that sort always contain a mixed chirality fermion pair and therefore do not 
interfere with SM $pp \to h\, W$ amplitudes. Here, $\Hdag \overleftrightarrow D^{\mu} H = (D^\mu\Hdag)\, H - \Hdag (D^{\mu} H)$.
 \end{itemize}

The left-hand column of Table ~\ref{contact_operators} shows the 12 operators derived from $D(Q^{\dag}Q\,\Hdag\, H W_{L,R})$, 
grouped into CP-even/odd pairs.  These operators are each accompanied by a real coefficient in the Lagrangian, however one 
could also combine each pair, e.g. $\mathcal O_{8,QW1}$ and $\mathcal O_{8,Q\tilde W1}$, into a complex operator with complex 
coefficient.  The 8 operators on the right-hand side correspond to $D(Q^\dag Q\ H^2 (\Hdag)^2)$ and $D^3(Q^\dag Q \Hdag H)$. 
These operators are accompanied by real coefficients.  Throughout this list we have chosen to put derivatives on the Higgs 
fields whenever possible. However, while it is possible to form an invariant with three derivatives on Higgs fields, 
$Q^{\dag}Q\,D^2\Hdag DH$ and $Q^{\dag}Q\,D\Hdag D^2H$ (each with two electroweak index contraction possibilities), the two 
are not independent under IBP so we cannot span the full set of $D^3(Q^{\dag}Q\Hdag H)$ operators with them. Rather than 
choose one, $Q^{\dag}Q\,D^2\Hdag DH$ {\em or} $Q^{\dag}Q\,D\Hdag D^2H$, we have opted for a more symmetric choice involving 
two derivatives on Higgs fields and one on a fermion field.

Using Table~\ref{contact_operators} one can easily write down similar operators involving right-handed fermions. For 
left-handed leptons we just need to replace $Q \rightarrow L$ since the $SU(3)_c$ structure played no role; similarly, 
trading in $Q$ for $u_c, d_c$ or $e_c$, only the $SU(2)_w$-singlet fermion combinations 
$\mathcal O_{8,Q1}, \mathcal O_{8,3Q1}, \mathcal O_{8,3Q3}$ and 
$\mathcal O_{8,QW1},\mathcal O_{8,Q\tilde W1}, \mathcal O_{8,QW2}, \mathcal O_{8, Q\tilde W2}$ are allowed.
   
\section{Application: $pp \to h\, W^{\pm}$}
\label{application}

To investigate the effect of dimension-8 operators we focus on one process: Higgs boson production in association with a 
$W$ boson at the LHC.  See Refs.\,\cite{Khachatryan:2016vau, Sirunyan:2017elk, Sirunyan:2018egh, Sirunyan:2018ouh, Aaboud:2017xsd, 
ATLAS:2016gld} for relevant experimental results. 

Combining the operators in Tables~\ref{17_derivative_free}-\ref{contact_operators} with dimension-6 operators (enumerated in 
Appendix~\ref{app:dim6}),  higher-dimensional terms manifest in a number of ways.  Bosonic operators directly enter into 
$pp \to h W^{\pm}$ by modifying the $hWW$ vertex, while fermionic operators (e.g. Table~\ref{contact_operators}) either modify 
the $\bar q q W^{\pm}$ vertex or enter as $\bar q q h W^{\pm}$ contact terms. Additionally, higher-dimensional operators 
introduce corrections to the SM field kinetic terms.  For instance, $\mathcal O_{8,HB}$ leads to a correction to the $U(1)$ 
kinetic term, $c_{8,HB} (v^4/\Lambda^4) B_{\mu\nu}B^{\mu\nu}$, and similarly $\mathcal O_{8,HD}$ leads to a correction to the 
Higgs kinetic term. To ensure that all fields are canonically normalized we must make a set of field redefinitions. These 
redefinitions lead to shifts in electroweak parameters like couplings and mass terms. One needs to make a choice of 
experimental input parameters and shift others as derived parameters. For a discussion of this procedure in the case of 
dimension-6 operators, see Refs.~\cite{Alloul:2013naa,Degrande:2016dqg}. Dimension-8 operators introduce a dependence of the 
electroweak parameters on the shifts at order $1/\Lambda^4$ and need to be handled with care.  Details of this procedure are 
presented in Appendix~\ref{app:redefs}. 

After carrying out the normalization and EW input procedure, we next sketch the Feynman rules for the $\bar q qW$, $hWW$, and 
$\bar q q h W$ vertices. We take all momenta to be ingoing and enforce on-shell conditions on the Higgs and fermion fields, 
but not on the gauge bosons. While it is possible to remove the dependence on one field momentum in each vertex by imposing 
momentum conservation, we choose not to do so. To be more compact and make the different Lorentz structures clearer, we first 
express the Feynman rules in terms of form factors $c_{ffVi}, c_{hWWi}, c_{ffWhi}$\footnote{For simplicity, we neglect 
CP-odd operators throughout this discussion.}:

\begin{align}
\begin{fmffile}{ffV}
\begin{tikzpicture}[baseline=(current bounding box.center)]
\node{
\fmfframe(1,1)(1,1){
\begin{fmfgraph*}(50,40)
\fmfleft{i1,i2}
\fmfright{f1}
\fmf{fermion}{i1,v1}
\fmf{fermion}{v1,i2}
\fmf{wiggly}{v1,f1}
\fmflabel{$u(p_1)$}{i1}
\fmflabel{$d(p_2)$}{i2}
\fmflabel{$W^+_\mu(p_3)$}{f1}
\end{fmfgraph*}
}
};
\end{tikzpicture}
\end{fmffile}
& \quad\quad\quad\quad =   \overline v(p_2) \gamma^{\mu} \Big( c_{qqV0} + c_{qqV1}\,\frac{p^2_3}{2} \Big) P_L\, u(p_1)
\label{eq:ffV}
\end{align}

\begin{align}
\begin{fmffile}{hWW}
\begin{tikzpicture}[baseline=(current bounding box.center)]
\node{
\fmfframe(1,1)(1,1){
\begin{fmfgraph*}(40,40)
\fmfleft{i1}
\fmfright{f1,f2}
\fmf{dashes}{i1,v1}
\fmf{wiggly}{v1,f1}
\fmf{wiggly}{v1,f2}
\fmflabel{$h (p_1)$}{i1}
\fmflabel{$W^+_\mu (p_2)$}{f2}
\fmflabel{$W^-_\nu (p_3)$}{f1}
\end{fmfgraph*}
}
};
\end{tikzpicture}
\end{fmffile}
& \quad\quad=  \Big( c_{hVV0}\, \eta^{\mu\nu} + c_{hVV1}\, ((p_2 \cdot p_3)\eta^{\mu\nu} - p^{\nu}_2\, p^{\mu}_3) + c_{hVV2} ((p_1\cdot p_3)\eta^{\mu\nu} - p^{\nu}_1p^{\mu}_3) \nonumber \\[-7mm]
& \quad\quad\quad\quad\quad\quad\quad\quad - c^*_{hVV2} ((p_1\cdot p_2)\eta^{\mu\nu} - p^{\mu}_1\,p^{\nu}_2) \Big)
\end{align}
\vspace{5mm}

\begin{align}
\begin{fmffile}{udhW}
\begin{tikzpicture}[baseline=(current bounding box.center)]
\node{
\fmfframe(1,1)(1,1){
\begin{fmfgraph*}(50,40)
\fmfleft{i1,i2}
\fmfright{f1,f2}
\fmf{fermion}{i1,v1}
\fmf{fermion}{v1,i2}
\fmf{dashes}{v1,f1}
\fmf{wiggly}{v1,f2}
\fmflabel{$u(p_1)$}{i1}
\fmflabel{$d(p_2)$}{i2}
\fmflabel{$h(p_3)$}{f1}
\fmflabel{$W^+_\mu(p_4)$}{f2}
\end{fmfgraph*}
}
};
\end{tikzpicture}
\end{fmffile}
& \quad\quad=   \overline v(p_2)\Big(\gamma^{\mu}(c_{qqWh0} + c_{qqWh2}\, (p_3\cdot p_4) + c_{qqWh3}\, (p_1 \cdot (p_3+ p_4)) + c_{qqWh4}\, (p_2 \cdot (p_3 + p_4)) \nonumber \\[-7mm]
& \quad\quad\quad\quad\quad\quad\quad\quad+ \slashed{p_3}(c_{qqWh1}\, p^{\mu}_3 + c_{qqWh3}\, p^{\mu}_1 + c_{qqWh4}\, p^{\mu}_2)   \\
& \quad\quad\quad\quad\quad\quad\quad\quad\quad + \slashed{p_4}(-c_{qqWh2}\, p^{\mu}_3 + c_{qqWh3}\, p^{\mu}_1+ c_{qqWh4}\, p^{\mu}_2) ) 
\Big)P_L\, u(p_2) \nonumber
\end{align}
The full expressions for the form factors are provided in Appendix~\ref{app:formfactor}\footnote{While we have presented 
off-shell vertices, it would be interesting to explore these results using on-shell amplitude techniques along the lines 
of\,\cite{Azatov:2016sqh}.}.

Using these vertices to calculate $\hat{\sigma}(pp \to h\,W^+)$ in terms of the dimension-6 or -8 coefficients, the full 
expression is not particularly illuminating. However,  a quick way to see how dimension-8 effects enter and what are the most 
important operators is to take the limit of large $\hat s$, as that will expose differences in the high energy behavior 
of dimension-6 vs. dimension-8. We find:
\begin{align}
& \hat{\sigma} ( p p \to W^+ h) \sim  \Big(\frac{\hat e^2}{4608\,\pi\,\sin^4{\hat{\theta}}}\Big)\frac{\hat v^2}{m^2_W}
  \frac{\hat s}{\Lambda^4}\Big( e^2\,(c_{8,3Q1} - c_{8,3Q2} + c_{8,3Q3} + c_{8,3Q4} ) \nonumber \\[-3mm]
& \quad\quad\quad\quad\quad\quad\quad\quad\quad\quad\quad\quad\quad\quad\quad\quad\quad\quad\quad\quad\quad\quad\quad +8\,\sin^2{\theta}\,(c^{(3)}_{Hq})^2 \Big)  + \mathcal O(\hat s^0)
\label{eq:larges}
\end{align}
The largest growth is linear in $\hat s$, as expected from simple power-counting arguments, and the operators that enter are 
a dimension-6 contact term (squared), and a combination of $D^3(Q^{\dag}Q\Hdag H)$ dimension-8 terms.  The fact that 
$D^3(Q^{\dag}Q\Hdag H)$ terms are the only dimension-8 terms to appear is not surprising. Fermionic operators with fewer 
derivatives contain additional Higgs fields and can only contribute to $pp \to h\,W^+$ if multiple Higgs fields are set to 
their vevs (more vevs in the amplitude lead to weaker energy dependence). The only dimension-6 term that contributes in 
Eq.~\eqref{eq:larges} is 
$\mathcal O^{(3)}_{Hq} = i\, (Q^{\dag}\bar{\sigma}^{\mu}\tau^I\, Q)\, (\overleftrightarrow D\Hdag\,\tau^I H)$. However, this 
operator also modifies fermion couplings to $W$ and $Z$ bosons (through the form factor $c_{ffV0}$ in Eq.~\eqref{eq:ffV}). 
Strong constraints on deviations of $W$ and $Z$ couplings to fermions implies that these operators must have small coefficients, 
see e.g. Ref.~\cite{Masso:2014xra}. Examining Eq.~\eqref{eq:ffV}, we see that the $D^3(Q^{\dag}Q\Hdag H)$ contact terms are not 
tied to $\bar q q V$ modifications. Thus, if we take $c^{(3)}_{Hq} \to 0$ to avoid $\bar q q V$ constraints, the part of 
$\hat{\sigma}(pp \to h\,W^+)$ that grows with energy is controlled by dimension-8 operators alone. In this case, the cross section 
contributions from $|A_\dsix|^2 $ are $ \propto \frac{v^4}{\Lambda^4}$, while those from $A_{\textrm{SM}} \times A_\deight$ are 
$\propto \frac{v^2 \hat s}{\Lambda^4}$ -- so there are energy regimes where dimension-8 effects are dominant at $1/\Lambda^4$.  To 
quantitatively evaluate the effects we can expect at the proton level, we turn to numerics.

\begin{figure}[h!]
\centering
\includegraphics[width=0.45\textwidth]{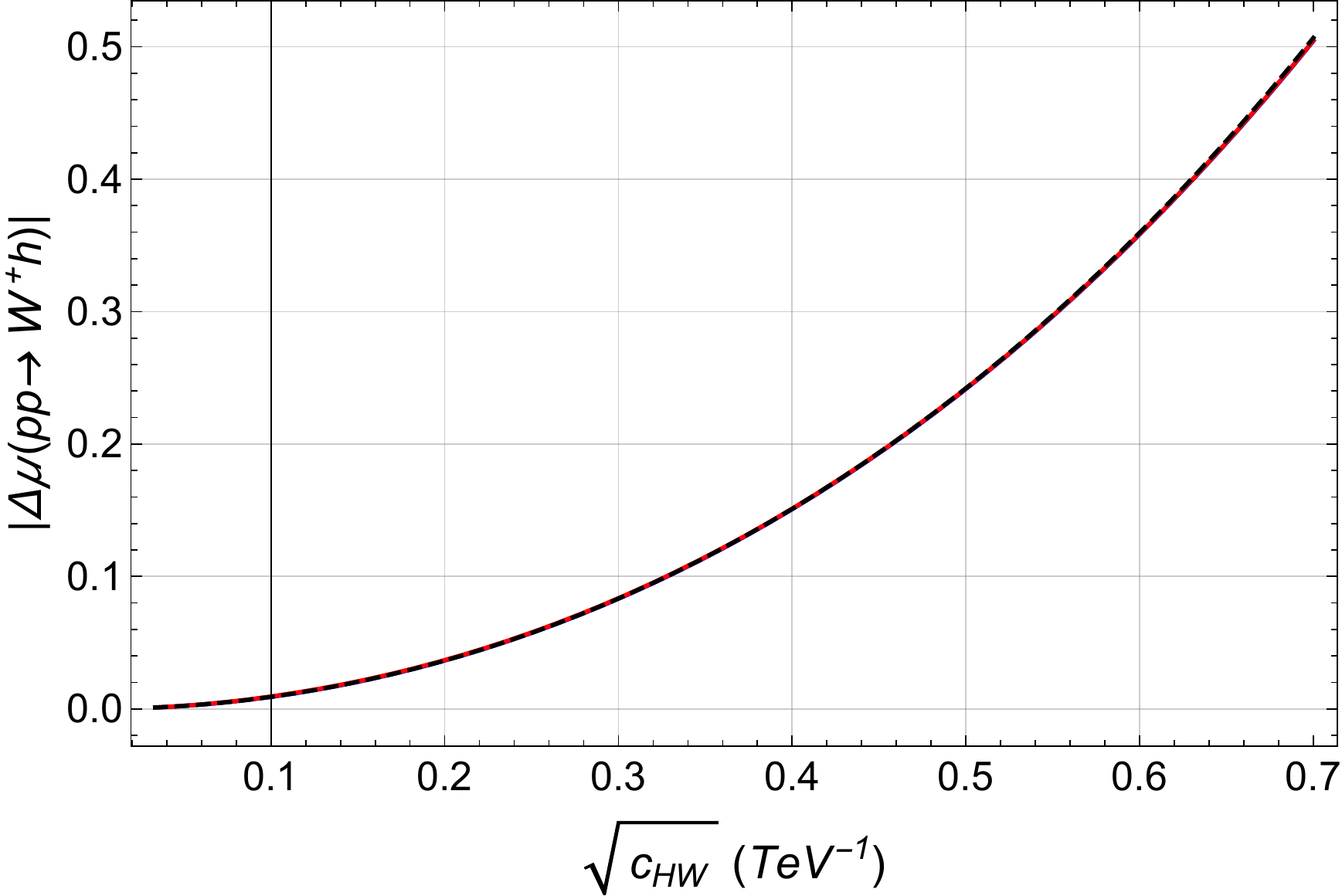}
\includegraphics[width=0.45\textwidth]{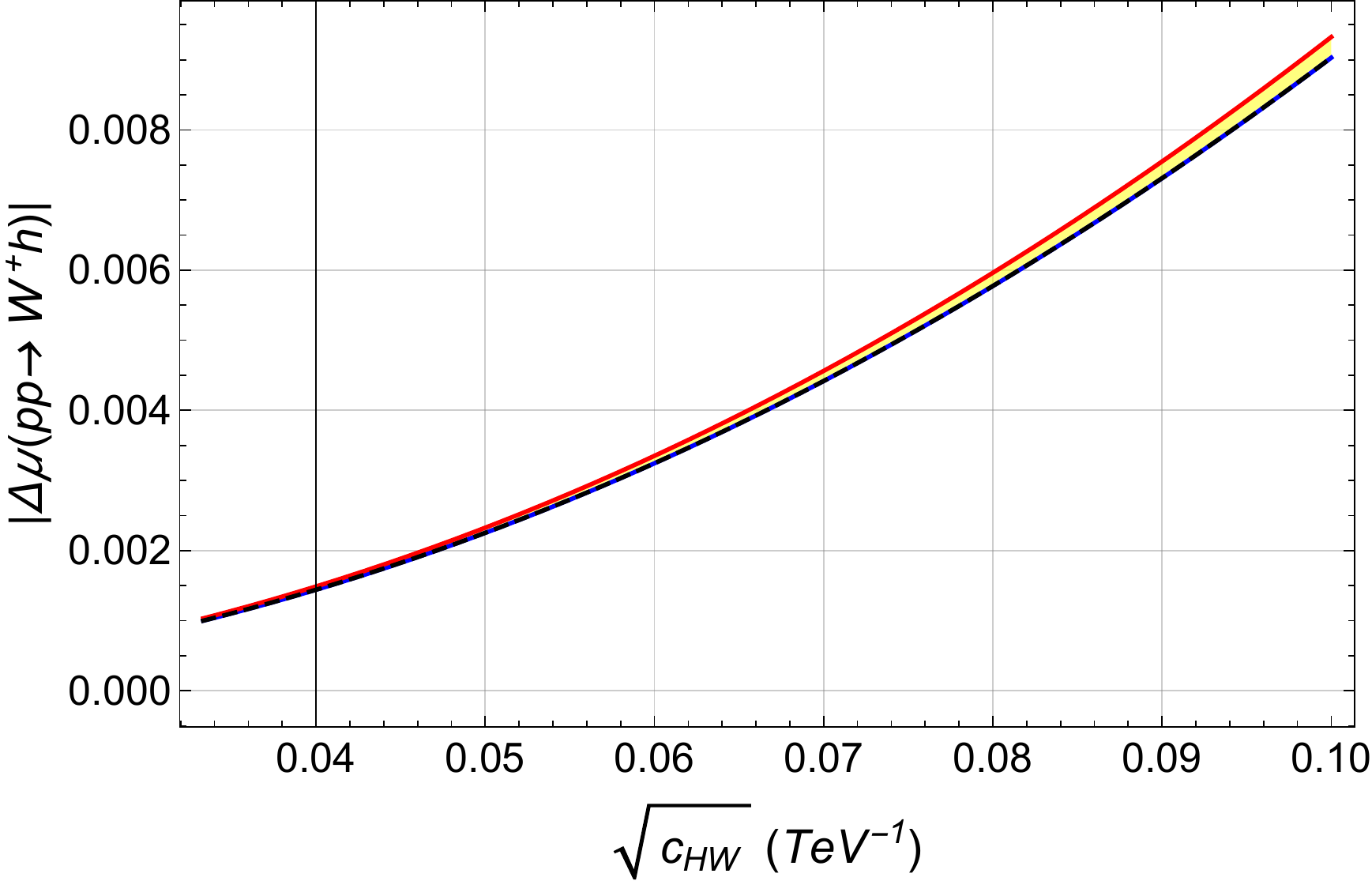} \\
\includegraphics[width=0.45\textwidth]{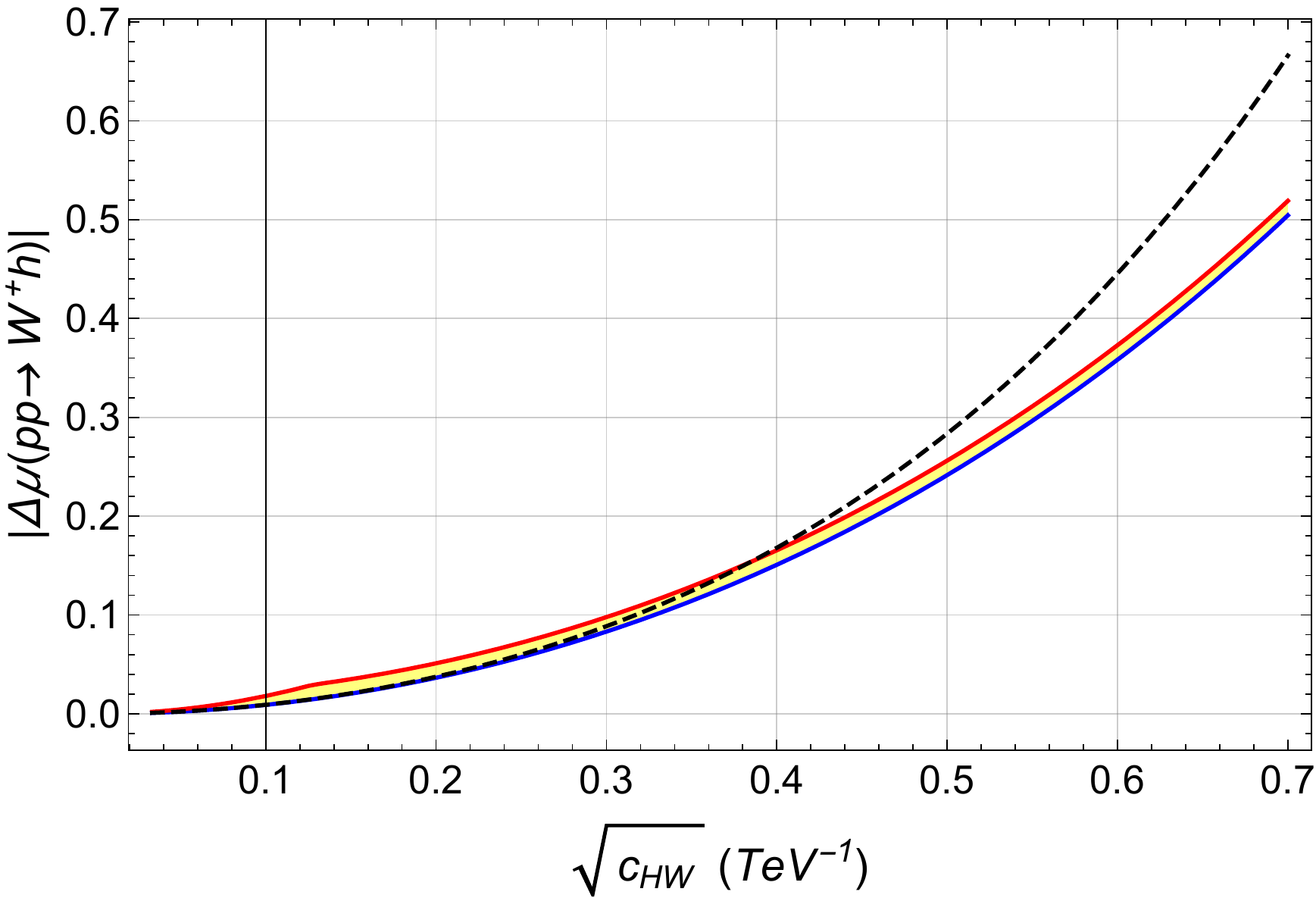}
\includegraphics[width=0.45\textwidth]{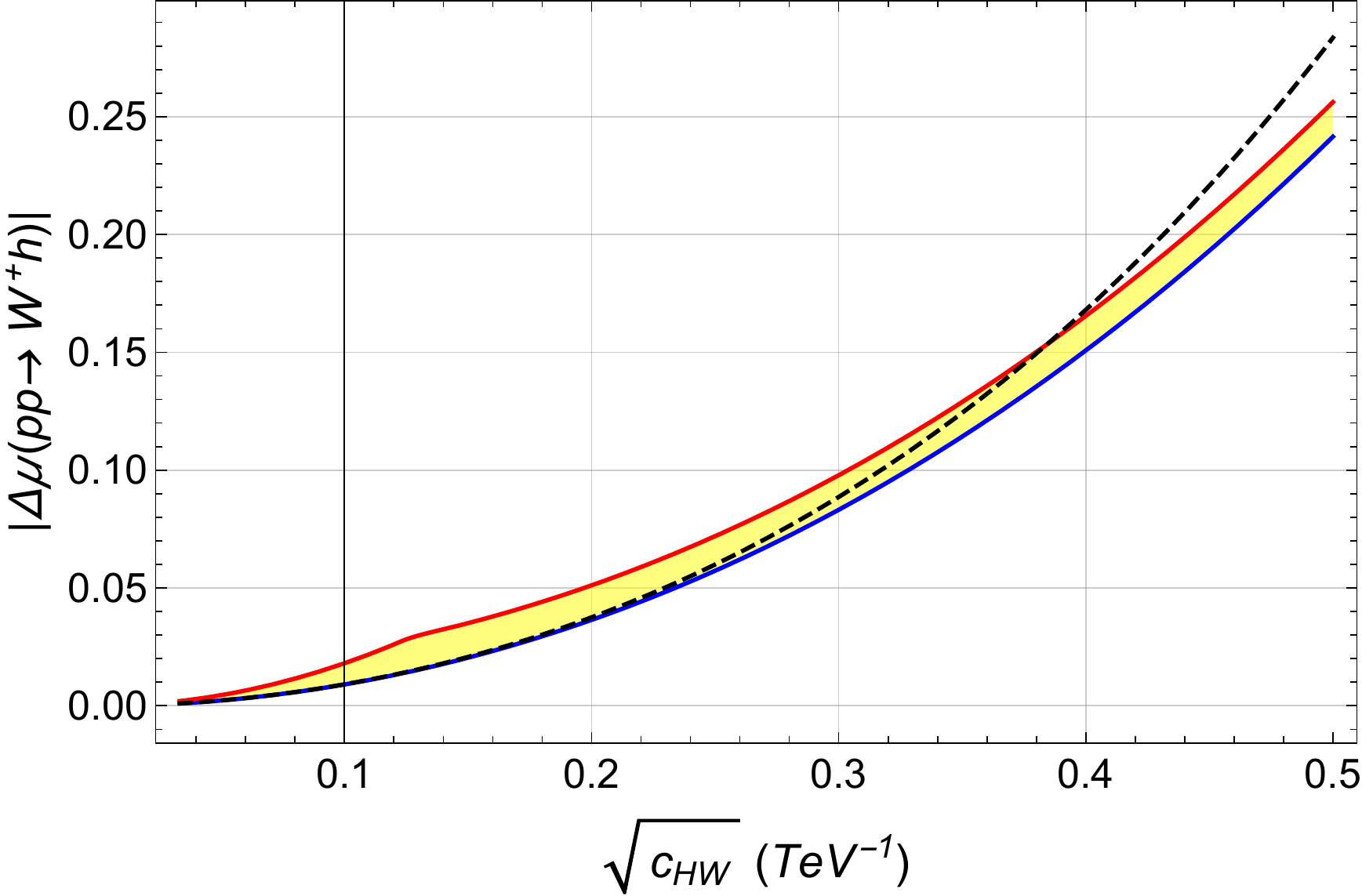}
\caption{Relative deviation in the inclusive cross section $\sigma(pp \to h\,W^+)$ from its SM value including dimension-6 and 
dimension-8 effects.  The blue line shows the result of including $\mathcal O_{HW}$ as the only dimension-6 operator and without 
considering dimension-8 operators.  The red line indicates the deviation as a function of $c_{HW}$ including the maximum possible 
dimension-8 effect consistent with the EFT expansion. The black dashed line shows the result if dimension-6 and dimension-8 
operator coefficients are equal, $c_{HW} = c_{8,i}$ (i.e. $\Lambda_8 = \Lambda_6$). In the top two panels the dimension-8 coefficients 
are all equal, while in the bottom two panels we take their magnitudes to be equal but assign their signs to maximize their effects at 
high $\sqrt{s}$. The left panels shows values of $\sqrt{c_{HW}}$ out to the current 95\% CL limit, which following the global analysis 
in Ref.~\cite{Ellis:2018gqa} is $0.631\, \tev^{-1}$.  In the right panels we have zoomed in to smaller values of $c_{HW}$ to make the 
dimension-8 effects more visible.}
\label{fig:ppWhin}
\end{figure}

As a rough estimate of the impact of dimension-8 operators, we study the rate of $pp \to h\,W^+$ in a scenario with a single 
dimension-6 operator and all dimension-8 operators.  We choose $\mathcal O_{HW}$ as the representative dimension-6 operator 
and for simplicity define $c_{HW} \equiv 1/\Lambda^2_6$ and set all other dimension-6 coefficients to zero.  We take all 
dimension-8 operator coefficients to have the same magnitude, $|c_{8,i}| \equiv 1/\Lambda^4_8$ but leave the signs to float 
since there can be cancellations among different operators (see Eq.~\eqref{eq:larges}). For a fixed $c_{HW}$ ($\Lambda_6$), 
the limit $\Lambda_8 \to \infty$ corresponds to no dimension-8 effects.  Decreasing $\Lambda_8$, we add in the dimension-8 
effects. For each $\Lambda_6, \Lambda_8$, and sign choice for the coefficients\footnote{As demonstrated in 
Refs.~\cite{Adams:2006sv,Bellazzini:2014waa}, it is possible that analyticity and unitarity requirements forbid certain 
signs for higher dimensional operator coefficients.  We ignore this possibility here and assume the coefficients can 
have either sign.}, we fold the parton-level results with parton distribution functions\footnote{We 
use MSTW2008nnlo~\cite{Martin:2009iq} parton distribution functions with factorization scale set to $\sqrt{\hat s}$.}, then 
calculate the shift in the $pp \to h\,W^+$ rate relative to the SM, 
$|\Delta \mu(pp \to h\,W^+)| = |(\sigma(pp \to h\,W^+)_{\Lambda_6, \Lambda_8} - \sigma(pp \to h\,W^+)_{SM})/\sigma(pp \to h\,W^+)_{SM}|$. 
If we pick the signs of all dimension-8 coefficients to be positive, the result is shown in the top panels Fig.~\ref{fig:ppWhin}. 
In the bottom panels of Fig.~\ref{fig:ppWhin} we show the result if we instead choose signs of the dimension-8 coefficients that 
enhance their contributions at large $\sqrt{s}$\footnote{The sign assignment is the following: 
$c_{8,HD2}, c_{8,HDHW}, c_{8,QW3}, c_{8,QW5}, c_{8,3Q1},c_{8,3Q3},c_{8,3Q4}$ positive ($ = +\frac 1 {\Lambda_8^4}$); and 
$c_{8,HD}, c_{8,HWB}, c_{8,HW}, c_{8,HW2}, c_{8,Q2},c_{8,Q3},c_{8,Q4}, c_{8,HDHW2}, c_{8,3Q2} $ negative.}.

The blue line in Fig.~\ref{fig:ppWhin} shows the relative deviation in $\sigma(pp \to h\,W^+)$ from the SM value as a function of 
$c_{HW}$, neglecting dimension-8 effects $(\Lambda_8 \to \infty)$. The impact of the dimension-8 operators can be seen by tracing 
either vertical or horizontal lines through Fig.~\ref{fig:ppWhin}.  Picking a value of $c_{HW}$ and tracing vertically upwards 
from the blue line intercepts two lines with different dimension-8 scenarios.  The black dashed line corresponds to the case 
where dimension-6 and dimension-8 operators have the same coefficient, $c_{HW} = c_{8,i}$ or $\Lambda_6 = \Lambda_8$.  The red 
line denotes where $\Lambda_8$ has been reduced to the point that the EFT breaks down, and thus represents the maximum potential 
dimension-8 contribution to $\sigma(pp \to h\,W^+)$.  This breakdown point occurs when one of the following validity conditions is 
broken: i) the $A_{\textrm{SM}} \times A_{\deight}$ contribution to the cross section of $\mathcal O(1/\Lambda^4_8)$ is greater than 
the quadratic dimension-8 contribution, $|A_{\deight}|^2 \sim \mathcal O(1/\Lambda^8_8)$; or ii.) the SM interference with dimension-6 
at $\mathcal O(1/\Lambda^2_6)$ is larger than SM interference with dimension-8 at $\mathcal O(1/\Lambda^4_8)$.  The first condition 
is independent of the dimension-6 effect, while the latter ties the two terms together.  As $c_{HW}$ decreases ($\Lambda_6$ 
increases), the second condition becomes the stronger of the two, causing the tapering in the band of dimension-8 effects.  We 
also require the condition $A_{\textrm{SM}}\times A_\dsix > |A_\dsix|^2$ in all calculations, which sets the maximum $c_{HW}$ value 
(the right-hand edge of the plot).

Comparing the top and bottom panels of Fig.~\ref{fig:ppWhin}, the band of dimension-8 effects is significantly smaller when we choose 
all dimension-8 coefficients to have the same sign. The origin of this difference is an accidental cancellation among the dimension-8 
terms when they have the same sign and magnitude. Specifically, the $c_{8,3Q1}$, $c_{8,3Q3},$ and $c_{8,3Q4}$ terms positively 
interfere with the SM amplitude, while the $c_{8,3Q2}$, $c_{8,QW3}$, and $c_{8,QW5}$ terms interfere negatively. The latter do not 
appear in Eq.~\eqref{eq:larges} as they enter the cross section at $O(\hat s^0)$, proportional to $\hat v^4/(m^2_W\, \Lambda^4)$.  
Cancellations between the $\mathcal  O(\hat s)$ and $\mathcal O(\hat s^0)$ pieces are possible since the inclusive production of 
$W\, h$ is dominantly near threshold, where $\hat s \sim v^2$ and thus the two terms are similarly sized. One may think that an 
accidental cancellation in $A_{\deight}$ can be compensated by lowering $\Lambda_8$. However, the $|A_{\deight}|^2$ contribution has 
no such cancellation, so it overwhelms the $A_{SM} \times A_{\deight}$ piece even at relatively high $\Lambda_8$. Thus, the net 
result of the cancellation in $A_{\deight}$ and our EFT consistency conditions is that the dimension-8 effects for the positive sign 
choice are reduced to a sliver. The cancellation in $A_{\deight}$ is broken if we relax the assumption of equal size coefficients or 
fixed signs. Turning on various combinations of dimension-8 couplings and adjusting their signs, we find that results similar in size 
and shape to the bottom panels in Fig.~\ref{fig:ppWhin} are far more common; therefore we will use these signs when 
quantifying the dimension-8 effects.  The net effect of the interference is positive for this choice but negative interference is also 
possible, in which case the yellow band would lie beneath the blue line. 
 
To get an estimate of how much dimension-8 operators affect the extraction of the dimension-6 coefficients, we return 
to Fig.~\ref{fig:ppWhin} and trace horizontally through a fixed value of $|\Delta \mu(pp \to h\,W^+)|$.  Let us take 
$|\Delta \mu (pp \to h\,W^+)| = 0.2$ as an example. Extending a horizontal line through that point in the bottom panels, 
we intersect the red line corresponding to the maximum considered dimension-8 effects at $\sqrt{c_{HW}} = 1/(2.19\,\text{TeV})$ 
and the blue line corresponding to no considered dimension-8 effects at $\sqrt{c_{HW}} = 1/(2.27\, \text{TeV})$.  
At this level of $|\Delta \mu (pp \to h\,W^+)|$, the relative difference is quite small, $\le 4\%$.  However, for 
$\Delta \mu (pp \to h\,W^+) = 0.05$, the relative size of the dimension-8 effects in our treatment is $\approx 18\%$, 
roughly spanning $\sqrt{c_{HW}} = 1/(4.28\,\text{TeV})$ for no dimension-8 effects to $\sqrt{c_{HW}} = 1/(5.08\,\text{TeV})$ 
including maximal dimension-8 effects. To see the impact of $\Lambda_8 = \Lambda_6$, rather than the maximal allowed 
dimension-8 effect, we repeat the above procedure but look for where a line of constant $|\Delta \mu(pp \to h\,W^+)|$ 
intersects the dashed black line. For $|\Delta \mu(pp \to Wh)|$ = 0.2, the intersection lies outside of the yellow band, 
meaning that the $\Lambda_8 = \Lambda_6$ point lies outside our definition of EFT validity. For $|\Delta \mu(pp \to h\,W^+)| = 0.05$, 
the intersection is at $\sqrt{c_{HW}} = 1/(4.36\, \text{TeV})$, a $2\%$ shift from the dimension-6 value.  For reference, 
ATLAS projects a $Wh(\to \bar{b}b)$ precision at the HL-LHC of $|\Delta \mu| \sim 0.14$~\cite{ATL-PHYS-PUB-2014-011}, and a 
global precision including all channels of $<10\%$~\cite{ATL-PHYS-PUB-2014-016}.  We emphasize that our results use a single 
(but representative) sign assignment and equal-magnitude dimension-8 coefficients, and are therefore only indicative.  
   
\begin{figure}[t!]
\centering
\includegraphics[width=0.45\textwidth]{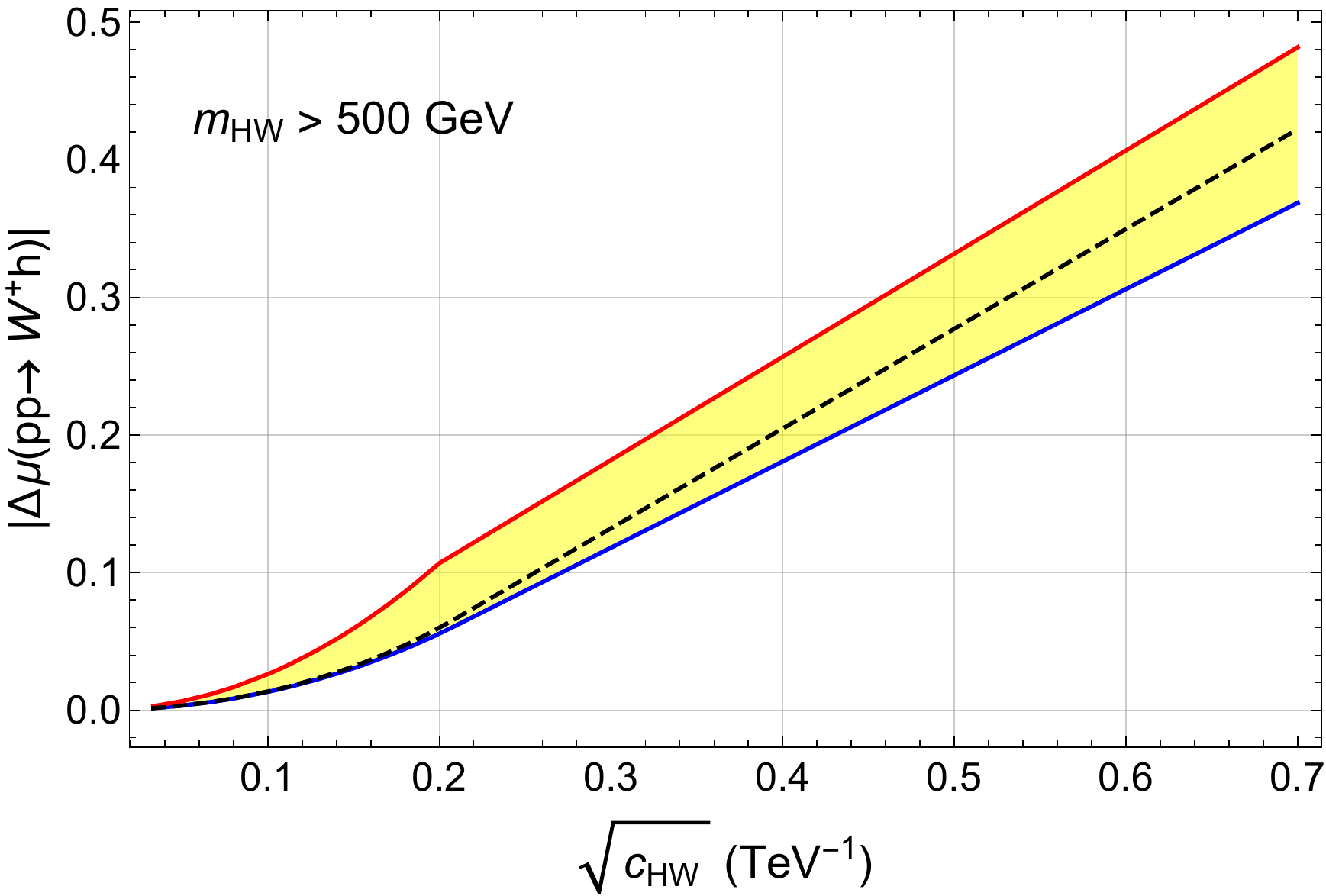} 
\includegraphics[width=0.45\textwidth]{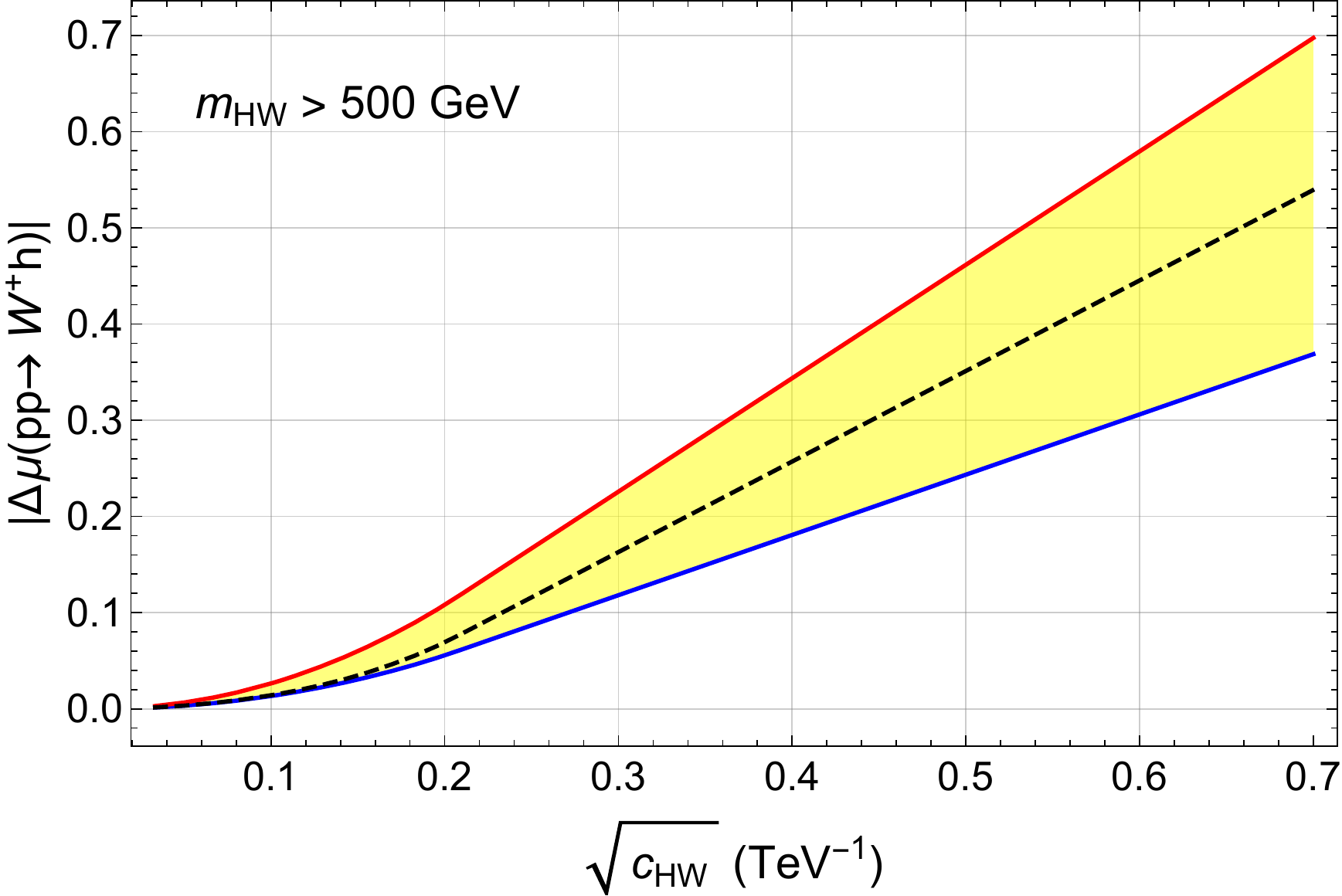}
\caption{
Relative deviation in the high-mass cross section $\sigma(pp \to h\,W^+, m_{HW} > 500\, \gev)$ from its SM value including 
dimension-6 and dimension-8 effects. In the left panel, all dimension-8 coefficients are taken to be positive, while in the 
right panel the signs of the coefficients enhance the impact of dimension-8 operators on this cross section.  The 
blue, red, and dashed black lines correspond to the same scenarios as in Fig.~\ref{fig:ppWhin}.}
\label{fig:ppWhhighmass}
\end{figure}

To see how the dimension-8 operators affect high-scale kinematic regions, we repeat the $\sigma(pp \to h\,W^+)$ calculation 
focussing on a region of high invariant mass, $m_{HW} \equiv \sqrt{\hat s}  > 500\, \gev$.  The results are shown in 
Fig.~\ref{fig:ppWhhighmass}, both for the case where all dimension-8 coefficients are positive and for the sign assignment in 
Fig.~\ref{fig:ppWhin}. Compared to the inclusive case, the effects of adding dimension-8 operators are significantly 
larger and the EFT validity conditions (which must be recalculated for this kinematic region) carve out a different shape.  The 
increased $\hat s$ also disrupts the cancellation in $A_{\deight}$ for the inclusive cross section when all dimension-8 
coefficients are taken to have the same sign.  Quantifying the effect, in the mixed-sign case a measurement of 
$|\Delta \mu(pp \to h\,W^+)|_{m_{HW} > 500\,\gev} = 0.2$ can be interpreted as 
$\sqrt{c_{HW}} = 1/(2.32\, \text{TeV})$ neglecting dimension-8 operators and $\sqrt{c_{HW}} = 1/(3.59\, \text{TeV})$ including 
maximal dimension-8 effects (a $\sim 55\%$ difference). For $\Lambda_8 = \Lambda_6$, the effect shrinks to 
$27\%$ ($\sqrt{c_{HW}} = 1/(2.95\, \text{TeV})$)\footnote{For the common-sign case, $\sqrt{c_{HW}} = 1/(3.08\, \text{TeV})$ 
including maximum dimension-8 effects, and $\sqrt{c_{HW}} = 1/(2.54\, \text{TeV})$ for $\Lambda_8 = \Lambda_6$.}.

The impact of dimension-8 would be significantly smaller had we neglected the contact terms, 
illustrating the importance of including all operators that can contribute to a process.  This statement is not intended 
to give the impression that contact terms are special, as the fact that they are the operators with contributions that 
grow with $\hat s$ is an artifact of our use of the Warsaw basis. In other bases, such as the SILH~\cite{Giudice:2007fh} 
basis, contributions growing with $\hat s$ would still be present though not necessarily originating from contact terms. The 
relative importance of the different operators would also be different in a scenario with unequal coefficients.

The trends exhibited in Figs.~\ref{fig:ppWhin} and~\ref{fig:ppWhhighmass} are not surprising: the more precisely a quantity 
is measured, the more sensitive it is to higher order corrections; and direct probes of high scales are more sensitive to 
higher-dimension operators.  However, this is the first time dimension-8 effects have been quantified in an LHC process 
using the complete set of dimension-8 operators.

Our analysis has assumed that only $c_{HW}$ is non-zero. This operator enters $\sigma(pp \to h\,W^+)$ at 
$\mathcal O(\hat s^0)$, whereas other dimension-6 contributions carry different energy dependence. We have seen that 
$c^{(3)}_{HQ}$ enters at $\mathcal O(\hat s)$, while other operators---in particular, operators that only contribute to 
$pp \to h\,W^+$ via normalization or electroweak inputs---enter at $\mathcal O(\hat s^{-1})$. As the current bounds on 
$c_{HW}, c^{(3)}_{HQ}$, etc. are not radically disparate, it is interesting to investigate the dimension-8 effects in 
scenarios with different dimension-6 energy dependence. The results of repeating the analysis in this section 
for $c^{(3)}_{HQ}$ or $c_{H\Box}$ variations can be found in Appendix~\ref{app:alternate}.

\section{Conclusions}

In this paper we have evaluated the effect of dimension-8 operators on two Higgs observables, in the context of the 
Standard Model EFT. For this purpose we have compiled a complete and non-redundant set of dimension-8 operators involving 
gauge bosons and at least one Higgs boson, along with the fermionic contact operators necessary to study $h\,W$ production 
at the LHC. A vital tool aiding in this construction was the Hilbert series, which tells us how many independent operators 
exist for each combination of fields, accounting for possible redundancies due to the equations of motion and integration 
by parts. Applied to the SMEFT, we find 17 independent operators with zero derivatives, 26 with two derivatives, 3 with 
four derivatives, and 20 operators involving $h$, $W$ and left-handed quarks. 

Through a series of examples, we outlined the steps required to convert between Hilbert series output---the number of 
invariants and their field and derivative content---to a canonical, phenomenology-ready form including Lorentz and gauge indices. 
This set of steps is completely general to relativistic EFTs with fields in linear representations of the defining symmetry 
groups, and is based on a method first proposed in Ref.~\cite{Lehman:2015coa}. The most involved step in the translation is 
the imposition of IBP redundancies, which is performed by constructing a matrix of IBP relations between the operators with 
all possible derivative partitions and the operators with the same field content but fewer derivatives, and then taking the 
matrix to row-reduced form. The translation procedure would clearly benefit from automation, especially if one would like to 
extend it to other sets of observables (see Ref.~\cite{Gripaios:2018zrz} for recent progress in that direction). We have made 
available the complete set of operators affecting $pp \to h\,W$ in FeynRules format in the hope that it will prove useful for 
future analyses.

We used this framework to study the impact of dimension-8 operators on the production of a Higgs boson in association with a 
$W^+$ boson. This channel provides a good handle on higher dimensional operators due to the kinematic reach of this topology. 
Higher dimensional operators contribute to the $\bar q q W$ vertex, $h W W$ vertex, and $\bar q q W h$ contact terms. 
These contributions accompany different energy dependencies in the cross section. In particular, several contact operators 
lead to $\sigma(pp \to h\,W^+)$ contributions that grow with $\hat s$. Unlike their counterparts at dimension-6, the 
dimension-8 operators that impact $\bar q q W h$ do not modify the trilinear $\bar q q V$ coupling.

To estimate of the effects of dimension-8 operators, we studied scenarios where only one dimension-6 operator coefficient 
is nonzero and all dimension-8 operator coefficients have equal magnitude but different sign.  We quantified the uncertainty 
on the extracted dimension-6 coefficient value by drawing contours of constant deviation in $\sigma(pp \to h\,W^+)$ and seeing 
where they intersected the predictions for additional dimension-6 terms and additional dimensions 6 and 8 terms.  Taking 
$c_{HW} \ne 0$ as the representative dimension-6 operator and varying the signs of the dimension-8 operator coefficients, 
we find the effects of dimension-8 are typically at the percent level for the inclusive cross section at 
its currently measured accuracy, growing to $\mathcal O(10\%)$ once we reach $|\Delta \mu(pp \to h\,W^+)| = 0.05$. The exact 
percentage varies depending on whether the dimension-8 effects act coherently or if there is some cancellation among different 
operators. We define the maximum size of the dimension-8 contribution by demanding EFT consistency, including the requirement 
that the contribution to the cross section linear in the dimension-8 coefficients makes a larger contribution to the cross section 
than the quadratic dimension-8 terms. If we focus on high-mass kinematic regions, the effects of dimension-8 operators become much 
larger. For example, for the coefficient set we use, dimension-8 operators shift the dimension-6 coefficient implied by 
$|\Delta\mu(pp \to h\,W^+)_{m_{HW} > 500\, \gev}| = 0.2$ by 55\%. In order to carry out these phenomenological studies, we 
performed canonical normalization and electroweak input procedures including dimension-8 effects (see Appendix~\ref{app:redefs}). 

This study and its companion implementation in FeynRules (see Appendix~\ref{app:FRs}) open the possibility for experimental 
collaborations and theoretical analyses to assign a systematic uncertainty to the effect of dimension-8 operators. Within 
the context of $pp \to h\, W^+$, it would be interesting to carry out uncertainty estimates for dimension-6 operators other 
than $c_{HW}, c^{(3)}_{HQ},$ or $c_{H\Box}$, or to more thoroughly explore the effects of correlations and cancellations among 
different operators. This work also allows theoretical studies of the interplay of dimension-8 operators with specific types 
of new physics, for example CP-odd Higgs couplings at dimension-6. 

Moving beyond $pp \to h\,W^+$, the logical next step is to extend the current FeynRules implementation to more operators and 
study the impact of dimension-8 effects on Higgs production in association with a $Z$ boson or with top quarks. It would also be 
interesting to add pure gauge dimension-8 operators (such as those in \cite{Degrande:2013kka}) which would affect di-boson and 
tri-boson production, and to relate these to the anomalous trilinear and quadrilinear gauge couplings (aTGC and aQGC), see 
for example references~\cite{wiki:CMS,Aaboud:2017tcq}. \\

{\it Note added:} As this paper was being completed, Ref.~\cite{Gripaios:2018zrz} appeared demonstrating a similar method for 
explicitly constructing non-redundant sets of higher dimensional operators in the SMEFT and beyond. That reference includes a 
software package automatizing the required steps.

\appendix

\section*{Acknowledgements}

The work of VS is supported by the Science Technology and Facilities Council (STFC) under grant number ST/P000819/1. The 
work of AM was partially supported by the National Science Foundation under Grant No. PHY-1520966. We would like to thank 
Andrea Banfi for discussions on the jet-veto effects in $Wh$ production. AM would like to thank Landon Lehman for his input during the initial stages of the project.

\section{Implementation in {\tt FeynRules}}
\label{app:FRs}

In this section we briefly discuss the implementation in FeynRules. Two {\tt .fr} files are included along 
with the source code of this paper, and will be available shortly in the {\it model database} in their 
\href{http://feynrules.irmp.ucl.ac.be}{webpage} (\url{http://feynrules.irmp.ucl.ac.be}). Both files contain 
the operators in Appendix~\ref{app:dim6} for dimension 6 and Tables~\ref{17_derivative_free}-\ref{contact_operators} 
for dimension 8. The operator coefficients in the {\tt .fr} files have the same name convention we have used in the text, e.g. 
{\tt c8HD} as the coefficient of $\mathcal O_{8,HD}$.  Within the files, the operators are grouped according to whether 
they influence $pp \to h\,W^+$, and the contact operators have been split into individual modules to speed up the code. The 
difference between the two files is whether or not the canonical normalization and electroweak input procedures described in 
Appendix~\ref{app:redefs} have been carried out. The procedures are needed to consistently include dimension-8 effects but 
they significantly slow down the running of the current incarnation of the {\tt .fr} files, so we have provided a version omitting 
that step.  We have made three other simplifications: 1.) we have omitted any four-fermion contributions to $G_F$, 2.) as our 
focus here is on $h\,W$ production we have neglected the $\mathcal O(\Lambda^{-4})$ difference between $\sin^2{\bar{\theta}}$ 
and $\sin^2{\theta_Z}$ in the coupling of $Z$ bosons to fermions, and 3.) we have omitted all CP-violating dimension-8 fermionic 
operators. We will address these shortcuts in future versions of the {\tt .fr} and plan to extend the set of translated 
operators to explore $pp \to Zh$ and $pp \to VV$. To ensure that events with these {\tt .fr} include dimension-8 interference 
effects but not $|A_\deight|^2$, one should generate events with the suffix {\tt NP$\string^$2 <=2}.

\section{Dimension-6 operators}
\label{app:dim6}

\begin{table}[h!]
\begin{center}
\begin{tabular}{|c|c|c|c|}
\hline
$\mathcal{O}_H $ & $ (H^\dagger H)^3 $ & 
$\mathcal{O}_{H\tilde{W}} $ & $ \delta_{IJ}(H^\dagger H)W^I_{\mu\nu}\widetilde{W}^J_{\mu\nu} $ \\
$\mathcal{O}_{H\Box} $ & $ (H^\dagger H) D^2 (H^\dagger H) $ &
$\mathcal{O}_{HB} $ & $ (H^\dagger H) B_{\mu\nu} B_{\mu\nu}  $ \\
$\mathcal{O}_{HD} $ & $ (D_\mu H^\dagger H)(H^\dagger D_\mu H) $ &
$\mathcal{O}_{H\tilde{B}} $ & $ (H^\dagger H) B_{\mu\nu} \widetilde{B}_{\mu\nu} $ \\
$\mathcal{O}_{HG} $ & $ \delta_{AB} (H^\dag H) G^A_{\mu\nu} G^B_{\mu\nu} $  &
$\mathcal{O}_{HWB} $ & $ \delta_{IJ} (H^\dagger \tau^I H) W^J_{\mu\nu} B_{\mu\nu} $ \\
$\mathcal{O}_{H\tilde{G}} $ & $ \delta_{AB} (H^\dag H) G^A_{\mu\nu} \widetilde{G}^B_{\mu\nu}  $ &
$\mathcal{O}_{H\tilde{W}B} $ & $ \delta_{IJ} (H^\dagger \tau^I H) \widetilde{W}^J_{\mu\nu} B_{\mu\nu} $ \\
$\mathcal{O}_{HW} $ & $ \delta_{IJ}(H^\dag H)W^I_{\mu\nu}W^J_{\mu\nu} $ & 
$\mathcal O^{(1)}_{HQ}$ & $i\, (Q^{\dag}\bar{\sigma}^{\mu}\,Q)(\Hdag \overleftrightarrow D^{\mu} H)$ \\
& & 
$\mathcal O^{(3)}_{HQ}$ & $i\, (Q^{\dag}\bar{\sigma}^{\mu}\tau^I\,Q)(\Hdag\,\overleftrightarrow D^{\mu}\tau^I\, H)$ \\
\hline
\end{tabular}
\caption{The thirteen dimension-6 operators included for comparison with dimension-8 effects. As described in the text, 
we work with the Warsaw basis. }
\label{11_dimension_6}
\end{center}
\end{table}

\section{Other examples}
\label{app:examples}

In this appendix we give two further examples to illustrate the conversion from the Hilbert series output to EFT operators 
in their canonical form.

\subsection{Example: $2\, (H^{\dag}H)^2\, W^2_L$}

In this example we will demonstrate the procedures for handling multiple operators and non-trivial $SU(2)_w$ contractions.  
The coefficient of the Hilbert series output for the operator $2\, (H^{\dag}H)^2\, W^2_L$ indicates that we need to find 
two invariants. The Higgs field portion of $(H^{\dag}H)^2\, W^2_L$ is identical to the example in Sec.~\ref{sec:example1}, 
so the Higgs group-theory decomposition is identical to Eq.~\ref{eq:H21}. The remaining object, $W_L$, is bosonic, so 
$(W_L)^2$ must be symmetric. However,  $W_L$ transforms under both $SU(2)_L$ and $SU(2)_w$ so there are more ways to form 
a symmetrized product. Specifically, $W^2_L$ can be overall symmetric if it is either symmetric in both $(SU(2)_L, SU(2)_w)$ 
indices {\em or} antisymmetric in both\footnote{Symmetrizing will depend on the representation we are working with; for 
$SU(2)$ triplets the symmetric combinations of $X_A\, Y_B$  (spin-0) are with $\delta^{AB}$ or the two-index symmetric 
tensor $X_{\{A}Y_{B\}}$ (spin-2), while the antisymmetric case is the antisymmetric tensor $X_{(A}Y_{B)} = $ vector, 
$\epsilon^{ABC} X_AY_B$ (spin-1). However if we deal with $SU(2)$ doublets, spin-0 is the antisymmetric combination 
$\epsilon^{ij}x_i y_j$ while spin-1 (vector) is the symmetric case $x_{\{i}y_{j\}}$.}:
\begin{align}
W^2_L &= (1,0;1)^2_{symm} = (0 \oplus 1 \oplus 2,0; 0 \oplus 1 \oplus 2)_{symm} = (0 \oplus 2,0;\, 0 \oplus 2) + (1,0;\, 1) \nonumber \\
 & = \Big( (0,0;0) + (2,0;0) + (0,0;2) + (2,0;2) \Big) + (1,0;\, 1).
\end{align}
We know the Higgs part of the operators is a Lorentz singlet, so only Lorentz-singlet $W^2_L$ options can make potential 
invariants:  $(0,0; 0)$ and $(0,0; 2)$. Adding in the Higgs fields,
\begin{align}
H^2 (H^{\dag})^2 (W^2_L) &= \Big((0,0;1)\otimes (0,0;1) \Big) \otimes (0,0;0\oplus 2) \nonumber \\
& = (0,0; 0 \oplus 2) \otimes (0,0; 0 \oplus 2), \nonumber
\end{align}
we can pick out the two invariants. One invariant comes from the product of the $SU(2)_w$ singlet element of $(\Hdag H)^2$ 
with the $SU(2)_w$ singlet piece of $W^2_L$, while the other comes from the $SU(2)_w$ spin-2 (symmetric tensor) piece of 
$(\Hdag H)^2$ with the corresponding piece of $W^2_L$.

For the product of $SU(2)_w$ singlets, the $H^2$ and $(H^{\dag})^2$ terms are contracted together\footnote{Which we can form 
either by inspection, or by taking the singlet product of the $H^2$ triplet and $(\Hdag)^2$ triplet, then simplifying via the 
identity $\tau^A_{ij}\tau^A_{lm} = 2\delta_{im}\delta_{jl} - \delta_{ij}\delta_{lm}$.}, as are the two $W_L$ fields, 
\begin{align}
(\epsilon \Hdag H)^2\, W^I_{L,\mu\nu}W^{I,\mu\nu}_L.
\label{eq:H4W21}
\end{align}
Similarly, to form the invariant from $SU(2)_w$ $2 \otimes 2$, we want to contract the symmetric product of the $H^2$, 
$(\Hdag)^2$ triplets with the symmetric product of $W_L$ triplets
\begin{align}
({\textrm{Tr}}(H\,\tau^A \eps H)\, {\textrm{Tr}}(\Hdag\,\tau^B \eps \Hdag) + A \leftrightarrow B)\, 
  (W^A_{L,\mu\nu}\, W^{B,\mu\nu}_{L} + A \leftrightarrow B),
\label{eq:H4W22}
\end{align}
where the $\epsilon$ in the $H^2$ product appears (as in Eq.~\eqref{eq:H4W21}) because $\Hdag$ is a $2$ of $SU(2)_w$. The two 
terms are identical, so the operator collapses to
\begin{align}
\Tr(H\,\tau^A \eps H)\, \Tr(\Hdag\,\tau^B \eps \Hdag)\, W^A_{L,\mu\nu}\, W^{B,\mu\nu}_{L}.
\end{align}
Technically, to form a true spin-2 representation from the product of spin-1 (triplet) representations, we should have subtracted 
a piece proportional to $\delta_{AB}$. However, as a $\delta_{AB}$ would reduce the operator in Eq.~\eqref{eq:H4W22} to the 
form Eq.~\eqref{eq:H4W21}, we can just absorb that contribution into the coefficient of Eq.~\eqref{eq:H4W21}. Repeating these 
steps for the hermitian conjugate term $(H^{\dag}H)^2\, W^2_R$ gets us terms analogous to Eqs.~\eqref{eq:H4W21} and~\eqref{eq:H4W22} 
but with $W_L \to W_R$. Finally, as with the example in Sec.~\ref{sec:example1}, we can separate the real and imaginary pieces 
into independent operators by considering linear combinations of the $W_L$ and $W_R$ forms. Written in terms of $W, \tilde W$ 
and with $\Hdag$ as a $\bar 2$, the four operators are:
\begin{align}
\mathcal O_{8,HW} = (\Hdag H)^2 W^I_{\mu\nu}W^{I,\mu\nu}, 
  & \quad \mathcal O_{8,H\tilde{W}} = (\Hdag H)^2 W^I_{\mu\nu}\tilde W^{I,\mu\nu} \\
\mathcal O_{8,HW2} =  (\Hdag\, \tau^I H)(\Hdag\, \tau^J\, H)\, W^I_{\mu\nu}W^{J,\mu\nu}, 
  & \quad \mathcal O_{8,H\tilde{W}2} = (\Hdag\, \tau^I H)(\Hdag\, \tau^J\, H)\, W^I_{\mu\nu}\tilde W^{J,\mu\nu}.
\end{align}

\subsection{Example: $4\,D(Q^{\dag}Q (\Hdag H)^2)$}

The goal of this example is to show how to manipulate operators with fermions and operators with a single derivative. As discussed in 
Sec.~\ref{sec:D0}, the first step is to enumerate the ways we can partition the derivative. The derivative cannot act on $Q$ or $Q^{\dag}$ 
since $DQ, DQ^{\dag}$ will transform under both Lorentz $SU(2)_L$ and $SU(2)_R$, while the other object in a non-trivial Lorentz 
representation only transforms under one of the two. The derivative can therefore either act on $H$ or $\Hdag$, so let us begin with $DH$. 
There are no repeated fermion fields so we do not need to worry about anytisymmetrization, and we can ignore $SU(3)$ since it is clear that 
we only want the color-singlet portion of $QQ^{\dag}$. Sticking with just Lorentz and $SU(2)_w$ indices and grouping terms conveniently:
\begin{align}
(Q^{\dag}Q)(H\, DH)(H^{\dag})^2 &= (\half, \half; 0 \oplus 1) \times (\half, \half; 0 \oplus 1) \times (0,0; 1) \nonumber \\
& =  (0,0; 0 \oplus 0 \oplus 1 \oplus 1 \oplus 1 \oplus 2) \times (0,0; 1).
\end{align}
This gives three terms, roughly: i.) the triplet of $Q^{\dag}Q$ contracted with the $\Hdag$ triplet, ii.) the triplet of $H\, DH$ contracted 
with the $\Hdag$ triplet, or iii.) contracting all three triplets with an $\eps^{IJK}$. Using $SU(2)_w$ indices the terms are:
\begin{align}
& i.)\,  (Q^{\dag}Q)_{\{ab\}} (H\, DH) (H^{\dag})^2_{\{cd\}}\eps^{ac}\eps^{bd} = 
  \Tr(Q^{\dag}\tau^I\, \eps\, Q) (H\, DH)\, \Tr(\Hdag\, \tau^J\, \eps\, \Hdag)\,\delta_{IJ}\nonumber \\
& ii.)\,(Q^{\dag}Q) (H\, DH)_{\{ab\}} (H^{\dag})^2_{\{cd\}}\eps^{ac}\eps^{bd} = 
  (Q^{\dag}Q)\, \Tr(H\,\tau^I\, \eps\, DH)\, \Tr(\Hdag\, \tau^J\, \eps\, \Hdag)\,\delta_{IJ} \nonumber \\
& iii.)\, (Q^{\dag}Q)_{\{ab\}} (H\, DH)_{\{ij\}} (H^{\dag})^2_{\{cd\}}\eps^{ai}\eps^{bc}\eps^{jd} = 
  \Tr(Q^{\dag}\tau^I\, \eps\, Q)\, \Tr(H\,\tau^J\, \eps\, DH)\, \Tr(\Hdag\, \tau^K\, \eps\, \Hdag)\eps_{IJK}. \nonumber 
\end{align}
Replacing $DH \to D\Hdag$ we have three more terms (iv., v., iv.), so a total of six. However, if we IBP on term ii.) above, we get
\begin{align}
& (Q^{\dag}Q) (H\, DH)_{\{ab\}} (H^{\dag})^2_{\{cd\}}\eps^{ac}\eps^{bd}  \to \nonumber \\
& (\text{total deriv.})- (Q^{\dag}Q) (DH\, H)_{\{ab\}} (H^{\dag})^2_{\{cd\}}\eps^{ac}\eps^{bd}  - 2(Q^{\dag}Q) (H^2)_{\{ab\}} (D\Hdag\, \Hdag)_{\{cd\}}\eps^{ac}\eps^{bd},
\end{align}
where we have ignored any terms with the derivative on the $Q, Q^{\dag}$ since we know they cannot yield invariants. Since the 
$SU(2)_w$ indices are all symmetric, the second term is the same as the original operator and the factor of 2 arises since 
differentiating either of the $\Hdag$ gives the same result. Ignoring the total derivative and rearranging, we find IBP gives:
\begin{align}
 (Q^{\dag}Q) (H\, DH)_{\{ab\}} (H^{\dag})^2_{\{cd\}}\eps^{ac}\eps^{bd} \to (Q^{\dag}Q) (H^2)_{\{ab\}} (\Hdag\, D\Hdag)_{\{cd\}}\eps^{ac}\eps^{bd}. 
\end{align}
This tells us that operators ii.) and v.) are not independent since we can always IBP on one to generate the other. The same 
manipulations work for operators iii.) and vi.), reducing the number of independent invariants to 4.  The same trick cannot be 
applied to operators i.) and iv.), since if we remove the derivative from $ (H\, DH) $ Bose symmetrization eliminates the 
operator\footnote{That is, we go from $\eps^{ij}H_i\, DH_j$ to $\eps^{ij}H_iH_j = 0$.}. The only surviving term in the IBP is 
when we shift the derivative from one $H$ to the other, getting us back operator i.). The 4 independent invariants are then
\begin{align}
& i.)\, \Tr(Q^{\dag}\bar{\sigma}^\mu\, \tau^I\, \eps\, Q) (H\, D_{\mu}\, H)\, \Tr(\Hdag\, \tau^J\, \eps\, \Hdag)\,\delta_{IJ}\nonumber \\
& ii.)\, \Tr(Q^{\dag}\bar{\sigma}^\mu\, \tau^I\, \eps\, Q) (\Hdag\, D_\mu \Hdag)\, \Tr(H\, \tau^J\, \eps\, H)\,\delta_{IJ}\nonumber \\
& iii.)\, (Q^{\dag}\bar{\sigma}^\mu\,Q)\, \Tr(H\,\tau^I\, \eps\, D_\mu H)\, \Tr(\Hdag\, \tau^J\, \eps\, \Hdag)\,\delta_{IJ}\quad \text{or}\quad DH \to D\Hdag \nonumber \\
& iv.)\, \Tr(Q^{\dag}\bar{\sigma}^\mu\, \tau^I\, \eps\, Q)\, \Tr(H\,\tau^J\, \eps\, D_\mu H)\, 
  \Tr(\Hdag\, \tau^K\, \eps\, \Hdag)\,\eps_{IJK}\quad \text{or}\quad DH \to D\Hdag \nonumber 
\end{align}
We can utilize Fierz rearrangement identities~\cite{Peskin:1995ev} to convert these into the forms shown in Table~\ref{contact_operators}.  

\section{Electroweak inputs and field redefinitions}
\label{app:redefs}

Expanded out to dimension-8, the electroweak sector of the SMEFT is a function of the gauge couplings, the Higgs quartic and vev, and the coefficients 
of the dimension-6 and -8 operators. In this appendix, we relate combinations of these inputs to precisely measured quantities.  The relationships are 
well known in the SM, and have been worked out previously for the dimension-6 SMEFT~\cite{Alonso:2013hga,Berthier:2015oma, Brivio:2017btx}.  The same 
methods are applied here, extended to include dimension-6-squared terms and linear dimension-8 terms as they are the same order in $1/\Lambda$.  As 
explained in the text, we work in the Warsaw basis throughout.

Only a subset of our operators are important for setting the EW inputs.  From dimension 6 they are 
$\mathcal O_H, \mathcal O_{H\Box}, \mathcal O_{HD}, \mathcal O_{HB}, \mathcal O_{HW}, \mathcal O_{HWB}$, while those from dimension-8 are 
$\mathcal O_{8,H}$, $\mathcal O_{8,HB}$, $\mathcal O_{8,HWB}$, $\mathcal O_{8,HW}$, $\mathcal O_{8,HW2}$, $\mathcal O_{8,HD}$, $\mathcal O_{8,HD2}$.  
In total, the EW sector at this order is a function of 17 inputs (13 operator coefficients, 2 gauge couplings, 1 quartic and 1 vev): 
\begin{align}
\label{eq:EWLAG}
\mathcal L_{EW,SM} &= -\frac 1 4 W^I_{\mu\nu}W^{I\mu\nu} - \frac 1 4 B_{\mu\nu}B^{\mu\nu} + (D_{\mu}\Hdag)(D^{\mu}H) - \lambda 
  \Big(\, \Hdag H - \frac{v^2_0}{2} \Big)^2, \nonumber \\
\mathcal L_{EW, 6} &=  \frac{c_H}{\Lambda^2} (\Hdag H)^3  + \frac{c_{H\Box}}{\Lambda^2}(\Hdag H)\Box(\Hdag H) + 
  \frac{c_{HD}}{\Lambda^2}((D_{\mu}\Hdag) H )(\Hdag\, D^{\mu}H) \nonumber \\ 
& \quad + \frac{c_{HW}}{\Lambda^2}(\Hdag H)W^I_{\mu\nu}W^{I\mu\nu}+ \frac{c_{HB}}{\Lambda^2}(\Hdag H)B_{\mu\nu}B^{\mu\nu} + 
  \frac{c_{HWB}}{\Lambda^2}(\Hdag\,\tau^I H)B_{\mu\nu}W^{I\mu\nu}, \nonumber \\ 
\mathcal L_{EW,8} &= \frac{c_{8,H}}{\Lambda^4}(\Hdag H)^4 + \frac{c_{8,HB}}{\Lambda^4}(\Hdag H)^2\,B_{\mu\nu}B^{\mu\nu} + 
  \frac{c_{8,HWB}}{\Lambda^4}(\Hdag H)(\Hdag\, \tau^I H)\,B_{\mu\nu}W^{I\mu\nu}  \nonumber \\
& \quad + \frac{c_{8,HW}}{\Lambda^4}\, (\Hdag H)^4W^I_{\mu\nu}W^{I\mu\nu} + 
  \frac{c_{8,HW2}}{\Lambda^4}(\Hdag \tau^I\, H)(\Hdag \tau^J\, H)\,W^I_{\mu\nu}W^{J\mu\nu} \nonumber \\
& \quad + \frac{c_{8,HD}}{\Lambda^4} (\Hdag H)^2(D_{\mu}\Hdag D^{\mu}H) + 
  \frac{c_{8,HD2}}{\Lambda^4} (\Hdag H)(\Hdag \tau^I H)(D_{\mu}\Hdag\,\tau^I D^{\mu}H).
\end{align}

We set the EW inputs using the $\{\alpha_{ew}, M^2_Z, G_F \}$ scheme, i.e. we solve for the gauge couplings and Higgs vev in terms of these 
observables and the coefficients of dimensions 6 and 8. To determine the Higgs quartic coupling we supplement our inputs with the measured 
Higgs mass, $M^2_H$. However, before we can relate the EW (and Higgs) inputs to observables calculated in the SMEFT theory, we need to bring 
Eq.~\eqref{eq:EWLAG} into canonical form.

The first step is to expand the Higgs field about its vacuum expectation value.  In the presence of higher dimensional operators, the minimum 
of the Higgs potential is no longer at $v_0$ but instead at
\begin{align}
\langle h \rangle = v_0\,\Big(1 + \frac{3\, c_H\, v^2_0 }{8\,\lambda\, \Lambda^2} + 
  \frac{v^4_0(63\, c^2_H + 32\, c_{8,H}\,\lambda)}{128\, \lambda^2\, \Lambda^4} \Big) \equiv v_T,
\end{align}
where $\lambda$ is the SM quartic. In unitary gauge the Higgs field is expanded as
\begin{align}
H = \frac 1 {\sqrt 2}\left(\begin{array}{c} 0 \\ (1 + c_{H,kin})h + v_T \\ \end{array} \right)
\end{align}
where $c_{H,kin}$ is the correction to canonically normalize the Higgs field, carried out to $\mathcal O(1/\Lambda^4)$: 
$c_{H,kin}=\frac{v^2_T}{4\,\Lambda^2}(4 c_{H\Box} - c_{HD}) + \frac{v^4_T}{32\,\Lambda^4}(3(c_{HD} - 4\,c_{H\Box})^2 -4\, c_{8,HD} - 4\,c_{8,HD2})$. 
Notice that the $1/\Lambda^{4}$ pieces of $c_{H,kin}$ and $v_T$ contain dimension-8 effects and effects from (dimension 6)$^2$. 

The next step is to canonically normalize the gauge fields: $B_{\mu} \to (1 + c_{B,kin}) \bar B_{\mu}, W^a_{\mu} \to (1 + c^a_{W,kin}) \bar W^a_{\mu}$ 
(barred fields are canonical). We can simultaneously redefine the gauge couplings to compensate for these changes, 
$g_1 \to \bar g_1/(1 + c_{B,kin}), g_2 \to \bar g_2/(1 + c_{W,kin})$, which has the effect that $g_1\, B_{\mu} = \bar g_1 \bar B_{\mu}$, etc., so 
that renormalizable gauge interactions in the dimension-8 SMEFT have the same form as the SM, but with barred couplings and fields. There is one 
subtlety here compared to dimension 6: the factor $c^a_{W,kin}$ is no longer universal for all $W^a$ as a consequence of $\mathcal O_{8,HW2}$.  We 
can only rescale $g_2$ once, so we must choose whether to absorb $c_{W^{\pm}, kin}$ or $c_{W^3,kin}$. Choosing 
$g_2 \to \bar g_2/(1 + c_{W^{\pm},kin})$, the neutral current at dimension-8 will no longer have the same form (in barred couplings and fields) as 
the SM~\cite{Grinstein:1991cd}. Explicitly:
\begin{eqnarray}
c_{W^{\pm}, kin} &= \frac{v^2_T}{\Lambda^2} c_{HW} + \frac{v^4_T}{2\Lambda^4}(3\, c^2_{HW} + c_{8,HW}) \nonumber \\
c_{W^{3}, kin} &= \frac{v^2_T}{\Lambda^2} c_{HW} + \frac{v^4_T}{2\Lambda^4}(3\, c^2_{HW} + c_{8,HW} + c_{8,HW2}) \\
c_{B, kin} &= \frac{v^2_T}{\Lambda^2} c_{HB} + \frac{v^4_T}{2\Lambda^4}(3\, c^2_{HB} + c_{8,HB}). \nonumber 
\end{eqnarray}

Next, we must diagonalize the kinetic and mass terms for the neutral gauge fields. This can be done following 
Refs.~\cite{Grinstein:1991cd, Alonso:2013hga}, 
\begin{align}
\left( \begin{array}{c} \bar W^3_{\mu} \\ \bar B_{\mu} \\ \end{array} \right) & =  
  \left( \begin{array}{cc} X_{11} & X_{12} \\ X_{12} & X_{11} \\ \end{array} \right)\left(\begin{array}{cc} \cos{\bar{\theta}} & \sin{\bar{\theta}} \\ 
 -\sin{\bar{\theta}} & \cos{\bar{\theta}} \\ \end{array} \right) \left(\begin{array}{c} \bar Z_{\mu} \\ \bar A_{\mu}\\ \end{array}\right) \nonumber \\
X_{11} &= 1 + \frac{v^4_T}{8\,\Lambda^4}(3\, c^2_{HWB}) \nonumber \\
X_{12} &= -\frac{v^2_T}{2\,\Lambda^2}c_{HWB} - \frac{v^4_T}{4\,\Lambda^4}(2\, c_{HWB} (c_{HB} + c_{HW}) + c_{8,HWB}), \nonumber 
\end{align}
where $\sin{\bar{\theta}}$ and $\cos{\bar{\theta}}$ are defined in terms of $v_T$, the barred couplings $\bar g_1, \bar g_2$ and other dimension-6 
and dimension-8 coefficients:
\begin{align}
\cos{\bar{\theta}} & = \frac{\bar g_2}{\sqrt{\bar g^2_1 + \bar g^2_2}} 
\Big(1 + \frac{v^2_T}{\Lambda^2}\frac{c_{HWB}}{2}\frac{\bar g_1}{\bar g_2}\frac{\bar g^2_1 - \bar g^2_2}{\bar g^2_1 + \bar g^2_2} + 
  \frac{v^4_T}{8\,\Lambda^4}\frac{\bar g_1\, }{\bar  g_2 (\bar g^2_2 + \bar g^2_1)^2} \times \nonumber \\
& \quad \quad \Big(2\, c_{8,HWB}(\bar g^4_1 - \bar g^4_2) + 4\, c_{8,HW2}\, \bar g_2 \bar g_1\, (\bar g^2_2 + \bar g^2_1) + 4\,c_{HWB}(c_{HW} + c_{HB})
  (\bar g^4_1 - \bar g^2_2) - \nonumber \\
& \quad \quad\quad \quad c^2_{HWB}\frac{\bar g_2}{\bar g_1}(\bar g^4_2 -6\, \bar g^2_1 \bar g^2_2 + 5\,\bar g^4_1) \Big)  \\
& \nonumber  \\
\sin{\bar{\theta}} &= \frac{\bar g_1}{\sqrt{\bar g^2_1 + \bar g^2_2}}\Big( 1 - \frac{v^2_T}{\Lambda^2}\frac{c_{HWB}}{2}\frac{\bar g_2}{\bar g_1}
  \frac{\bar g^2_1 - \bar g^2_2}{\bar g^2_1 + \bar g^2_2} +\frac{v^4_T}{8\,\Lambda^4}
  \frac{\bar g_2}{\bar g_1 (\bar g^2_1 + \bar g^2_2)^2} \times \nonumber \\
& \quad\quad \Big( 2\, c_{8,HWB}(\bar g^4_2 - \bar g^4_1) - 4\, c_{8,H\tilde{B}}\bar g_1\bar g_2 (\bar g^2_1 + \bar g^2_2) + 4\, c_{HWB}(c_{HW}+c_{HB})
  (\bar g^4_2 - \bar g^4_1) - \nonumber \\
& \quad\quad\quad\quad c^2_{HWB}\frac{\bar g_1}{\bar g_2}(\bar g^4_1 - 6\, \bar g^2_1\bar g^2_2 + 5\,\bar g^5_2) \Big).
\end{align}
From the diagonal form, we can read off the gauge-boson masses:
\begin{align}
\label{eq:mZ}
m^2_W &= \frac{\bar g^2_2\, v^2_T}{4} + \frac{\bar g^2_2\, v^6_T}{16\,\Lambda^4}(c_{8,HD} - c_{8,HD2} ) \\
m^2_Z &= \frac{v^2_T(\bar g^2_1 + \bar g^2_2)}{4} + \frac{v^2_T}{8\,\Lambda^2}\Big( c_{HD}(\bar g^2_1 + \bar g^2_2) + 4\,c_{HWB}\,\bar g_2 \bar g_1  \Big) + \\
& \frac{v^6_T}{16\,\Lambda^4} \Big((\bar g^2_1 + \bar g^2_2)(c_{8,HD} + c_{8,HD2} + 4\,c^2_{HWB}) + 
 \bar 4 g_1\bar g_2\,(c_{8,HWB} + c_{HWB}( 2\, c_{HB} + 2\, c_{HW} + c_{HD}) ) \nonumber \\
& \quad\quad\quad\quad\quad\quad\quad\quad\quad\quad\quad\quad\quad\quad\quad\quad\quad\quad\quad\quad+ 4\,\bar g^2_2\,c_{8,HW2} \Big). \nonumber 
\end{align}
\vspace{-0.1cm}
Expanding the covariant derivative and going to diagonal form, we can extract the couplings to the photon, $Z$ and $W$ bosons:
\begin{align}
D_{\mu}  = \partial_{\mu} + i\,\frac{\bar g_2}{\sqrt 2}(\bar W^+_{\mu}\tau^+ + \bar W^-_{\mu}\tau^- ) + 
  i\, \mathcal Q\, \bar e\, \bar A_{\mu} + i\, \bar g_Z (\tau_3 - \sin^2{\theta_Z}\,\mathcal Q)\,\bar Z_{\mu},
\end{align}
where:
\begin{align}
\label{eq:ebar}
& \bar e = \bar g_2\,\Big(\frac{ 1+ c_{W^{\pm}, kin}} {1 + c_{W^3, kin}} \Big)\,(\cos{\bar{\theta}}\, X_{12} + \sin{\bar{\theta}}X_{11})  \\
& \bar g_Z = \bar g_2\,\Big(\frac{ 1+ c_{W^{\pm}, kin}} {1 + c_{W^3, kin}} \Big)\, \frac{\rm{det}(X)}{\cos{\bar{\theta}}\, X_{22} + 
 \sin{\bar{\theta}}\, X_{21}} \nonumber \\
& \sin^2{\theta_Z}  =  - \frac{(\cos{\bar{\theta}} X_{12} + \sin{\bar{\theta}}\, X_{11})(\cos{\bar{\theta}} X_{21} 
  - \sin{\bar{\theta}}\, X_{22})}{\rm{det}(X)} \nonumber.
\end{align}
The expanded couplings are:
\begin{align}
& \bar e = \bar g_2\,\sin{\bar{\theta}} - \frac{v^2_T\,\bar g_2}{2\Lambda^2}\,c_{HWB}\,\cos{\bar{\theta}} - \nonumber \\
 & \, \, \, \, \, \, \, \frac{v^4_T\,\bar g_2\,}{8\Lambda^4}\Big(2\,( c_{8,HWB} + 2\, c_{HWB}(c_{HW} + c_{HB}))\cos{\bar{\theta}} - 
 (4\, c_{8,HW2} + 3\, c^2_{HWB})\sin{\bar{\theta}} \Big) \nonumber \\
& \bar g_Z = \frac{\bar g_2}{\cos{\bar{\theta}}} + \frac{v^2_T\,\bar g_2}{2\,\Lambda^2\, \cos^2{\bar{\theta}}} + \nonumber \\ 
& \, \, \, \, \, \, \,  \frac{v^4_T\,\bar g_2}{8\,\Lambda^4\, \cos^3{\bar{\theta}}}\Big(2\,\sin{\bar{\theta}}\cos{\bar{\theta}}(c_{8,HWB} + 
 2\, c_{HWB}(c_{HW} + c_{HB}) + \cos^2{\bar{\theta}}\, (c^2_{HWB} + 4\,c_{8,HW2}) \Big) \nonumber \\
& \sin^2{\theta_Z} = \sin^2{\bar{\theta}} + \frac{v^4_T}{4\,\Lambda^4}\,c^2_{HWB}(\sin^2{\bar{\theta}} - \cos^2{\bar{\theta}}).
\end{align}
The $(1 + c_{W^{\pm},kin})/(1 + c_{W^3, kin})$ factor in Eq.~\eqref{eq:ebar} is due to the different normalizations of $W^{\pm}$ and $W^3$ and 
is $\propto c_{8,HW2}$. As a result, $\sin^2{\theta_Z} \ne \sin^2{\bar{\theta}}$, meaning that the angle that rotates the gauge fields to mass 
eigenstates differs from the angle in the covariant derivative by $\mathcal O(1/\Lambda^4)$.\footnote{The usual technique for coding 
kinetic terms into FeynRules assumes $\sin^2{\theta_Z} \equiv \sin^2{\bar{\theta}}$. One quick fix to compensate for the mismatch at dimension-8 
is to include new operators, e.g. $\bar f\,\gamma^{\mu} f Z_{\mu}$ with coefficient $\propto (\sin^2{\theta_Z}-\sin^2{\bar{\theta}}$).}

Lastly, we express the Higgs boson mass as:
\begin{align}
M^2_{H} &= 2\lambda\, v^2_T\,\Big( 1 - \frac{v^2_T}{2\lambda\Lambda^2}(3\, c_H + \lambda (c_{HD} - 2\, c_{H\Box})) - 
  \frac{v^4_T}{4\,\lambda\,\Lambda^4}\Big( 3\,c_H(2\,c_{H\Box} - c_{HD}) \nonumber \\
& \quad\quad + 6\, c_{8,H} + \lambda\,(c_{8,HD} + c_{8,HD2} - (c_{HD} - 2\, c_{H\Box})^2) \Big) \Big).
\label{eq:mH}
\end{align}

We are now ready to set the EW inputs. Following~\cite{Alonso:2013hga}, it is convenient to write hatted quantities to represent those that are 
measured.  Using $\{\hat{\alpha}_{em}, \hat M^2_Z, \hat G_F, \hat M^2_H\}$, we can form the combinations:
\begin{align}
\label{eq:SMin}
\hat e = \sqrt{4\pi\hat{\alpha}_{em}},\quad \hat v^2 =&  \frac 1 {\sqrt 2 \hat G_F},\quad \sin^2{\hat{\theta}} = \frac 1 2 \Big( 1 - \sqrt{1 - \frac{4\pi\hat{\alpha}_{em}}{\sqrt 2\, \hat G_F\, \hat M^2_Z}}\, \Big), \\
\quad \hat g_1 = \frac{\hat e}{c_{\hatt}},&\quad \hat g_2 = \frac{\hat e}{s_{\hatt}},\quad \hat \lambda = \frac{\hat M^2_H}{2\,\hat v^2} \nonumber.
\end{align}
The task is now to solve for the barred input variables in terms of the hatted measured quantities, e.g. $\bar g_1(\hat{\alpha}_{em}, \hat M^2_Z, \hat G_F)$, or $\bar g_1(\hat e, \hatt, \hat v)$, with $\bar g_1 \to \hat g_1$ in the limit that all higher dimension coefficients vanish.

The Fermi constant $\hat{G}_F$ is set by muon decay and determines $\hat{v}$ through Eq.~\eqref{eq:SMin}. In the SM, muon decay comes 
from $W$-boson exchange, so $\hat{G}_F$ is the ratio of the charged-current (coupling)$^2$ divided by the $W$ boson mass.  The effects of 
higher dimensional operators are: i.) universal shifts to the charged-current coupling or $W$-boson mass, 
ii.) flavor-specific shifts in the charged current (e.g. shifts in $W$ boson coupling to $e\,\nu_e$ or $\mu\,\nu_{\mu}$), and  
iii.) 4-fermion contact terms. Calculated within the dimension-8 SMEFT, we find:
\begin{align}
\frac{\hat G_F}{\sqrt 2} \approxtext{dim-8}& \frac{\bar g^2_2}{8\, m^2_{W}} + \frac{\delta G_{F1}}{\Lambda^2} 
  + \frac{v^2_T\,\delta G_{F2}}{\Lambda^4} \nonumber \\
  = &\frac{1}{2\, v^2_T} - \frac{v^2_T}{4\,\sqrt 2 \Lambda^4}\Big(c_{8,HD} - c_{8,HD2} \Big) + \frac{\sqrt 2\, \delta G_{F1}}{\Lambda^2} 
  + \frac{\sqrt 2\, v^2_T\,\delta G_{F2}}{\Lambda^4},
 \label{eq:GF}
\end{align}
where we use $\delta G_{F1}, \delta G_{F2}$ to parametrize the non-universal contributions from fermionic dimension-6 and dimension-8 
operators, such as $\mathcal O^{(3)}_{H\ell}$ (in the notation of Ref.~\cite{Grzadkowski:2010es}), and the analog of 
Table~\ref{contact_operators} operators with $Q \to L$. Inverting Eq.~\eqref{eq:GF} defines $v_T(\hat v, c_i)$:
\begin{equation}
v_T = \hat v \Big(1 + \frac{\hat v^2}{\Lambda^2}\delta G_{F1} + \frac{\hat v^4}{8\,\Lambda^4}(-c_{8,H} + c_{8,HD2} 
  + 12\, \delta G^2_{F1} + 8\, \delta G_{F2}) \Big).
\end{equation}
Throughout the text, we have neglected leptonic operators that would contribute to $\delta G_{F1}, \delta G_{F2}$. Sticking strictly 
to the operators of our focus, one should set $\delta G_{F1}, \delta G_{F2} \to 0$, though we will maintain the dependence on 
$\delta G_{F1}, \delta G_{F2}$ in the following expressions.

Having solved for $v_T$, we can set Eqs.~\eqref{eq:ebar},~\eqref{eq:mZ} equal to the measured values $\hat e, \hat M^2_Z$, and then 
invert them to solve for $\bar g_1, \bar g_2$ (or some combination of them, such as $\sin{\bar{\theta}}, \cos{\bar{\theta}}$). Employing 
shorthand $\sin{\hatt} = s_{\hatt},\cos{\hatt} = c_{\hatt}$, etc:
\begin{eqnarray}
\bar g_1 &=& \frac{\hat e}{c_{\hat{\theta}}} \Big(1  + \frac{\hat v^2\, s_{\hat{\theta}}\,(4\, c_{HWB}\,c_{\hat{\theta}}+(c_{HD} + 
  4\, \delta G_{F1})\,s_{\hat{\theta}})}{4\,\Lambda ^2\,c_{2\hat{\theta}}} - 
  \frac{\hat v^4}{32\,\Lambda^4\, c^3_{2\hat{\theta}}} \times  \Big(-8\, c_{8,HD2}\,s^2_{\hatt}c^2_{2\hatt}   \nonumber \\
 & &\quad\quad -8\, c_{8,HWB}\,s_{2\hatt}c^2_{2\hatt} - 4\, c_{HWB}(c_{HW} + c_{HB})\,\frac{s^2_{4\hatt}}{s^2_{2\hatt}} 
 - 4\, c^2_{HWB}(6\,c_{2\hat{\theta}} + 3\,c_{4\hatt}+7)\, s^2_{\hatt} \nonumber \\
& &\quad\quad   + 8\, c_{HWB}\,c_{HD}(c_{2\hatt}-2)\,s_{\hatt}c^3_{\hatt} - c^2_{HD}(5\, c_{2\hatt}+2)s^4_{\hatt} -2\,\delta G_{F1}\,c_{HWB}(11\, s_{2\hatt} + 2\,s_{4\hatt} + 3\,s_{6\hatt})  \nonumber \\
& &\quad\quad -2\, \delta G_{F1}(2\,\delta G_{F1} +  c_{HD})(6\,c_{2\hat{\theta}}+ 3\,c_{4\hatt}+7)\,s^2_{\hatt} - 32\,\delta G_{F2}\,s^2_{\hatt}c^2_{2\hatt}\Big)\Big)  \\
 \bar g_2 &=& \frac{\hat e}{s_{\hat{\theta}}} \Big(1  - \frac{\hat v^2 c_{\hat{\theta}}  (4\, c_{HWB} s_{\hat{\theta}} + 
 (c_{HD} + 4\,\delta G_{F1})\,c_{\hat{\theta}})}{4\,\Lambda ^2\,c_{2\hat{\theta}}} - 
 \frac{\hat v^4}{32\,\Lambda^4\, c^3_{2\hat{\theta}}} \times  \Big(8\,c_{8,HD2}\,c^2_{\hatt}c^2_{2\hatt} + 16\,c_{8,HW2}\,c^3_{2\hatt}  \nonumber \\
& &  \quad\quad +8\,c_{8,HWB}\, s_{2\hatt}c^2_{2\hatt}+ 4\, c_{HWB}\,(c_{HB} + c_{HW}))\frac{s^2_{4\hatt}}{s_{2\hatt}} + 4\,c^2_{HWB}(-6\,c_{2\hatt} - 3\, c_{4\hatt}+ 7)\,c^2_{\hatt} \nonumber \\
& & \quad\quad + 8\,c_{HWB}c_{HD}(c_{2\hatt} + 2)\,s^3_{\hatt}c_{\hatt} - c^2_{HD}(5\, c_{2\hatt}-2)\,c^4_{\hatt} + 2\,\delta G_{F1}\,c_{HWB}(11\, s_{2\hatt} - 2\,s_{4\hatt} + 3\,s_{6\hatt})  \nonumber \\
& &\quad\quad  + 2\,\delta G_{F1}(2\,\delta G_{F1} +  c_{HD})(-6\,c_{2\hat{\theta}}-3\,c_{4\hatt}+7)\,c^2_{\hatt} + 32\,\delta G_{F2}\,c^2_{\hatt}c^2_{2\hatt} \Big)\Big) 
\end{eqnarray}

Lastly, we can set Eq.~\eqref{eq:mH} equal to the measured Higgs mass $\hat M^2_H$ and invert to solve for the quartic coupling.
\begin{eqnarray}
\lambda &=& \frac{\hat M^2_H}{2\, \hat v^2} +  \frac 1 {4\,\Lambda^2} (\hat M^2_H ( c_{HD} -4\,c_{H\Box} - 4\,\delta G_{F1}) + 6\, c_H\, \hat v^2) 
 + \frac{\hat v^2}{4\,\Lambda^4}\Big(\hat M^2_H\,(c_{8,HD} - 4\,\delta G_{F2}) \nonumber \\
 & &\quad\quad\quad\quad\quad\quad\quad \quad+ 6\,(c_{8,H}\, + 2\, c_H\, \delta G_{F1})\,\hat v^2 \Big)
\end{eqnarray}

With $\bar g_1, \bar g_2, v_T$ set, we can derive all other (EW) phenomenologically necessary parameters such as $m_W$ and $\sin^2{\theta_Z}$. 

\section{Explicit form for $\bar q q W, hWW$ and $\bar q q W h$ form factors}
\label{app:formfactor}

The form factors for the $\bar q q W, h\,WW$ and $\bar q q W h$ vertices are listed below. In the following we have performed the field and 
coupling redefinitions following Appendix~\ref{app:redefs}, but we have not re-expressed the couplings and vev in terms of measured EW inputs 
since that makes the expressions unwieldy.
\begin{eqnarray}
c_{qqV0} &=& \frac{i\, \bar g_2}{\sqrt 2} \Big( 1 + \frac{v^2_T}{\Lambda^2}\,c^{(3)}_{HQ} - \frac{v^4_T}{4\, \Lambda^4}(2\,c_{8,Q2} + 
 c_{8,Q4} - i\, c_{8,Q3})\Big) \nonumber \\
c_{qqV1} &=& \frac{i\, \bar g_2\, v^2_T}{2\,\sqrt 2 \Lambda^4}\,(c_{8,3Q2} - c_{8,3Q4}) \\
& \nonumber \\
c_{hVV0} &=& \frac{i\, \bar g^2_2\, v_T}{2}\Big( 1 + \frac{v^2_T}{4\Lambda^2}(4\,c_{H\Box} - c_{HD}) +  \frac{v^4_T}{8\,\Lambda^4}(5\,c_{8,HD} - 
  7\,c_{8,HD2} + \frac 3 4\,(c_{HD} -4\,c_{H\Box})^2 )\Big) \nonumber \\
c_{hVV1}  &=& i\, \Big( -\frac{4\, v_T}{\Lambda^2} c_{HW} - \frac{v^3_T}{\Lambda^4} (4\,c_{8,HW} - 8\, c^2_{HW} -c_{HW}(4\,c_{H\Box} - c_{HD})) \Big) 
\nonumber \\
c_{hVV2} & =& \frac{-i\, \bar g_2\, v^3_T}{4\Lambda^4}(c_{8,HDHW} - i\,c_{8,HDHW2}) \\
& \nonumber \\
c_{ffWh0} &=& i\,\sqrt 2\,\bar g_2\, v_T\,\Big( \frac{c^{(3)}_{HQ}}{\Lambda^2} - \frac{v^2_T}{4\,\Lambda^4}(4\, c_{8,Q2} + 2\, c_{8,Q4} - 
  2\, i\, c_{8,Q3} - c^{(3)}_{HQ}(4\, c_{H\Box} - c_{HD}) ) \Big) \nonumber \\
c_{ffWh1} &=& -i\, \frac{\bar g_2\, v_T}{\sqrt 2\, \Lambda^4 }\Big( c_{8,3Q1}+ c_{8,3Q3} \Big) \nonumber \\
c_{ffWh2} &=& -i\frac{\sqrt 2\,\bar g_2\, v_T}{\Lambda^4}(c_{8,QW3} + i\, c_{8,QW5})\nonumber \\
c_{ffWh3} &=& i\frac{\bar g_2\, v_T}{\sqrt 2\,\Lambda^4}(c_{8,3Q2})\nonumber \\
c_{ffWh4} &=& -i\frac{\bar g_2\, v_T}{\sqrt 2\,\Lambda^4}(c_{8,3Q4})
\end{eqnarray}

\section{Dimension-8 effects on other operators: $c^{(3)}_{HQ}$ and $c_{H\Box}$}
\label{app:alternate}

In this appendix we repeat the analysis of Sec.~\ref{application} in two other scenarios, one where the only non-zero dimension-6 operator 
is $\mathcal O^{(3)}_{HQ}$, and one with $\mathcal O_{H\Box}$ only. In both cases,  we set the dimension-8 operator coefficients using the 
mixed-sign configuration of Sec.~\ref{application} (i.e. all coefficients with equal magnitude, sign chosen to maximize effects at large 
$\sqrt{s}$). Expanding the partonic cross section $\hat{\sigma}(pp \to h\,W^+)$ for large $\hat s$, these dimension-6 operators contribute 
to different powers of $\hat s$ than $\mathcal O_{HW}$, so we expect the relative dimension-8 effects to differ from $\mathcal O_{HW}$.

\begin{figure}[t!]
\centering
\includegraphics[width=0.45\textwidth]{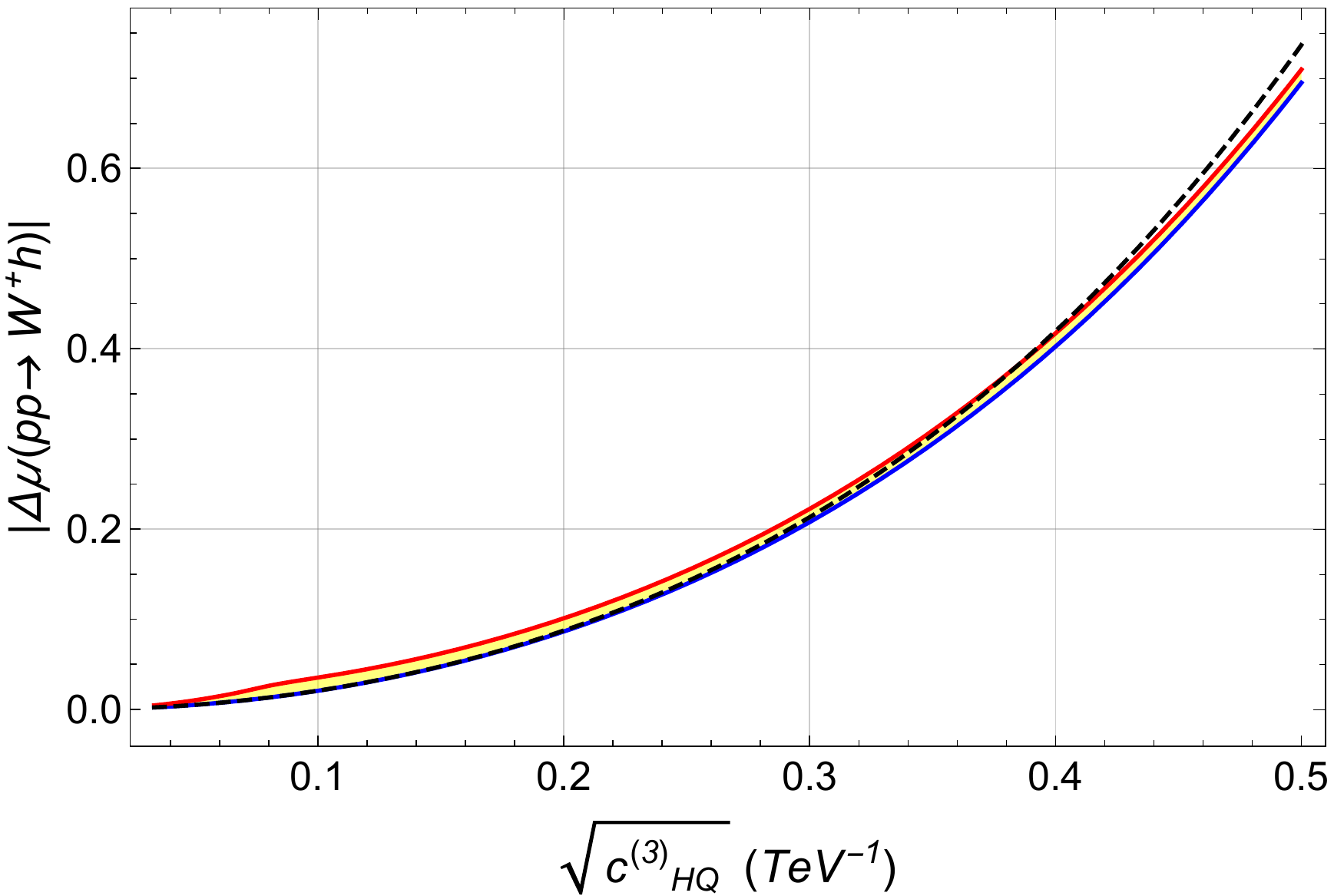}
\includegraphics[width=0.45\textwidth]{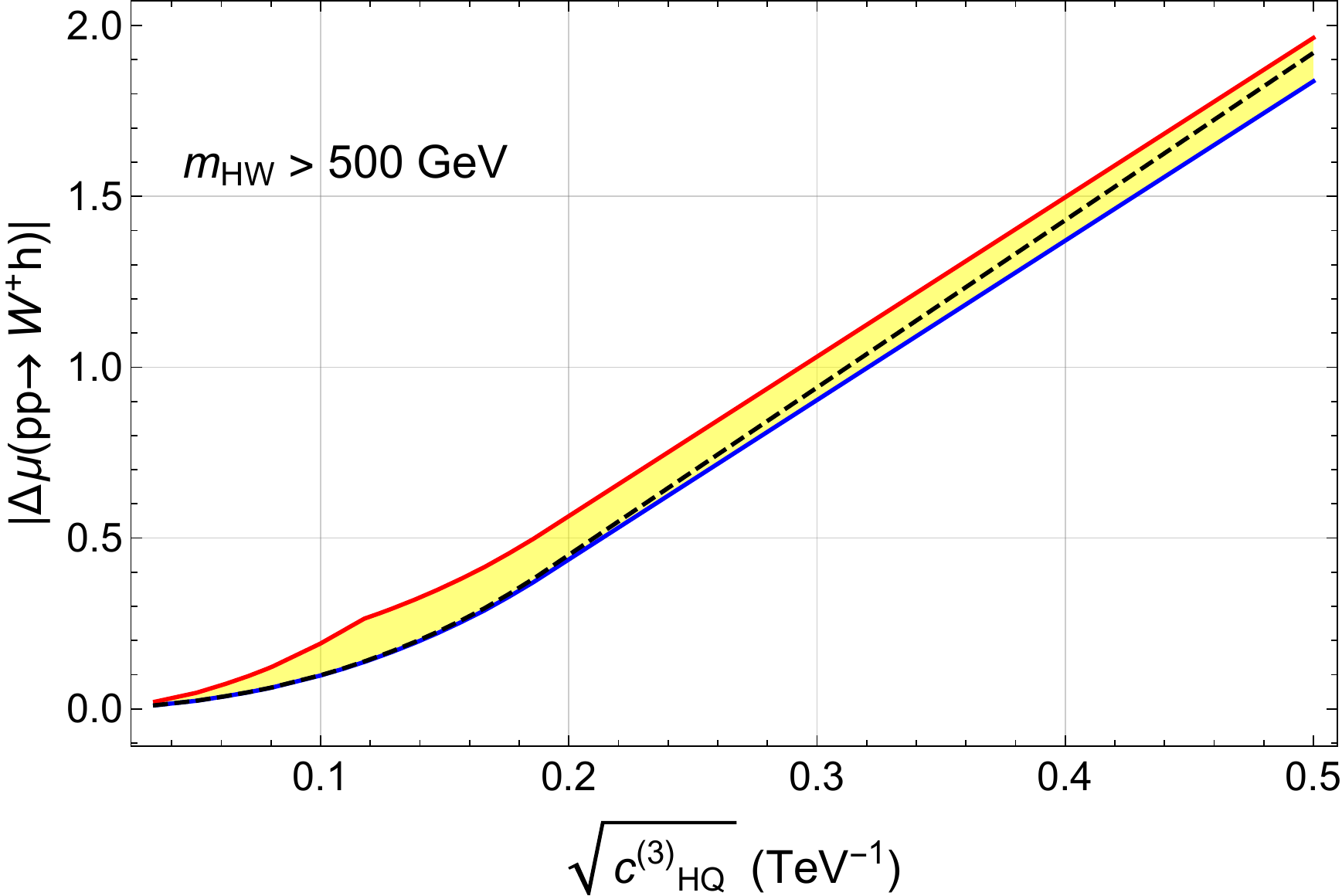}
\caption{Deviation in the inclusive (left panel) and high-mass (right panel) $\sigma(pp \to h\,W^+)$ cross section assuming the only non-zero 
dimension-6 operator is $\mathcal O^{(3)}_{HQ}$ and adding in all dimension-8 operators with equal magnitude coefficients and mixed signs as in 
the bottom panel of Fig.~\ref{fig:ppWhin}. The blue, red, and dashed black lines correspond to the same scenarios as in Fig.~\ref{fig:ppWhin}. 
The current limit on $c^{(3)}_{HQ}$ at 95\% CL is $0.66\,\tev^{-1}$~\cite{Ellis:2018gqa}; we have zoomed in to make the dimension-8 contribution 
more visible.} 
\label{fig:cHQ3}
\end{figure}

First we show the results for $\mathcal O^{(3)}_{HQ}$, whose coefficient $c^{(3)}_{HQ}$ modifies $\hat{\sigma}(pp \to h\,W^+)$ at 
$\mathcal O(\hat s)$.   Comparing the domains of Fig.~\ref{fig:cHQ3} with those of Figs.~\ref{fig:ppWhin}~and~\ref{fig:ppWhhighmass} we 
see that $c^{(3)}_{HQ}$ has a larger impact on $\sigma(pp \to h\,W^+)$ than $c_{HW}$. This is not a surprise given that $c^{(3)}_{HQ}$ 
produces stronger $\hat s$ dependence.  The dimension-8 effects are still present. For example, at $|\Delta \mu(pp \to h\,W^+)| = 0.2$, 
the variation between the scales inferred by the dimension-6 only interpretation and the dimension-6 plus maximum dimension-8 
interpretation is 3\% ($\Lambda_{NP} = 3.39\, \tev$ to $\Lambda_{NP} = 3.51\,\tev$), while for 
$|\Delta \mu(pp \to h\,W^+)|_{m_{HW} > 500\, \gev} = 0.2$ the difference in scales increases to 37\% ($\Lambda_{NP}= 7.12\,\tev$ to 
$\Lambda_6 = 9.79\, \tev$). In both panels, the tapering effect at low values of the dimension-6 coupling is less pronounced than it was 
in Figs .~\ref{fig:ppWhin} and~\ref{fig:ppWhhighmass}. This is due to the fact that a larger dimension-6 contribution to the cross section 
means $\Lambda_8$ can be lower before the $A_{SM}\times A_{\dsix} > A_{SM}\times A_{\deight}$ EFT validity criteria is violated. The 
$\Lambda_8 = \Lambda_6$ line in this scenario is difficult to see because it hugs the blue line.

\begin{figure}[t!]
\centering
\includegraphics[width=0.45\textwidth]{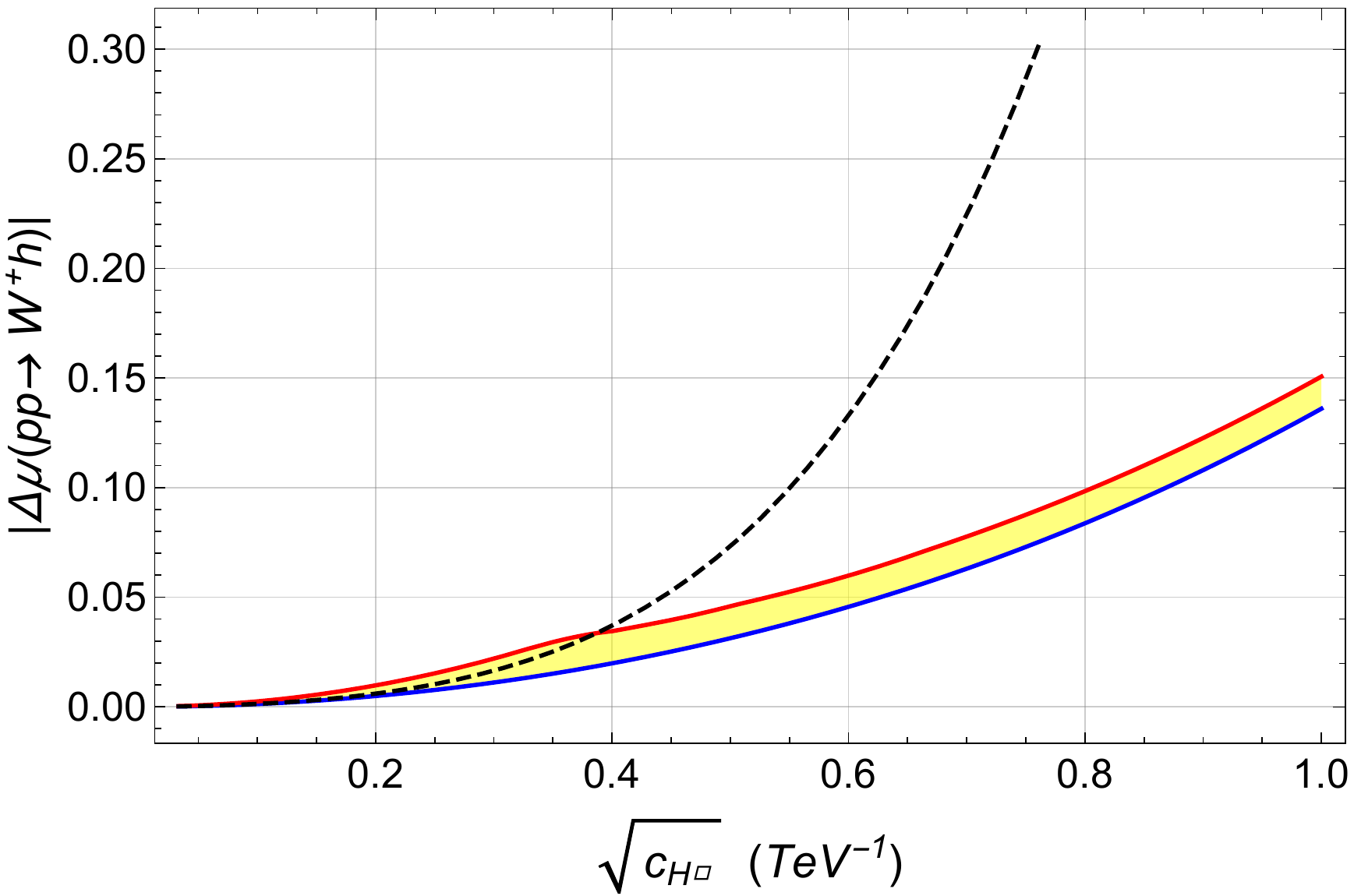}
\includegraphics[width=0.45\textwidth]{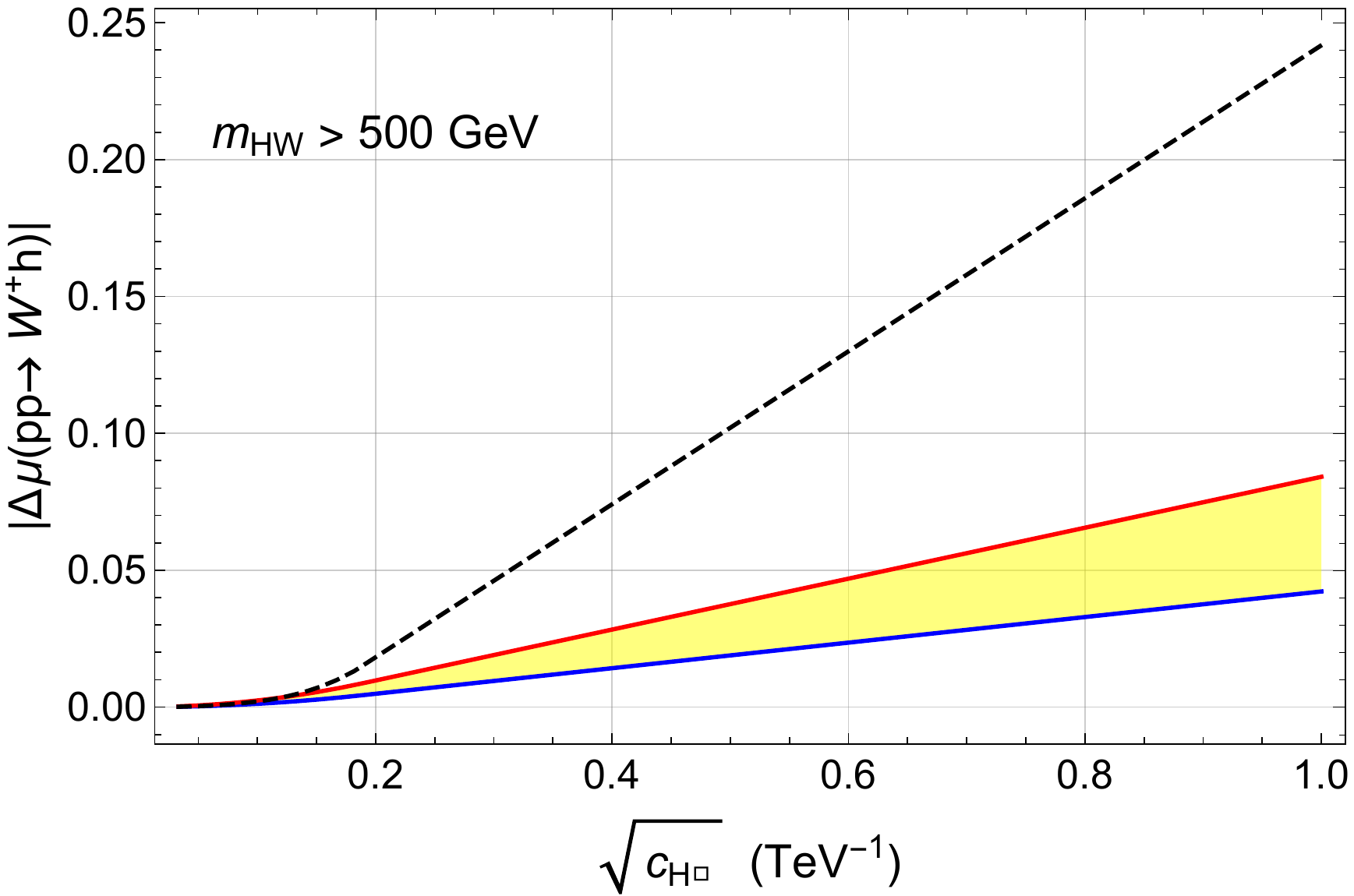}
\caption{Deviation in the inclusive (left panel) and high-mass (right panel) $\sigma(pp \to h\,W^+)$ cross section assuming the only non-zero 
dimension-6 operator is $\mathcal O_{H\Box}$ and adding in all dimension-8 operators with equal magnitude coefficients and the mixed signs as in 
the bottom panel of Fig.~\ref{fig:ppWhin}. The blue, red, and dashed black lines correspond to the same scenarios as in Fig.~\ref{fig:ppWhin}. 
Current constraints on $c_{H\Box}$ are weak, so we have zoomed in to make the dimension-8 contribution more visible.}
\label{fig:cHBox}
\end{figure}
Next, we repeat the exercise assuming the only dimension-6 operator is $\mathcal O_{H\Box}$, with coefficient $c_{H\Box}$.  
As $c_{H\Box}$ only modifies terms in the cross section at $\mathcal O(\hat s^{-1})$, its impact on the cross section is very small.  
One might expect that a suppressed dimension-6 piece could receive large relative dimension-8 corrections. However, our EFT validity 
requirement $A_{SM} \times A_{\dsix} > A_{SM} \times A_{\deight}$ prevents this from happening. Notice that the majority of the 
$\Lambda_6 = \Lambda_8$ line lies outside of the region where we trust the EFT. In terms of new physics scales inferred by measurements 
of $|\Delta\mu|$,  $|\Delta \mu(pp \to h\,W^+)| = 0.1$, the new physics scale ranges from $1.15\, \tev$ (no $\deight)$ to $1.24\, \tev$ 
(max $\deight)$, a 7\% shift.
  
\bibliography{references}
\bibliographystyle{JHEP}

\end{document}